\PassOptionsToPackage{table}{xcolor}

\documentclass[sigconf]{acmart}
\AtBeginDocument{%
  }

\setcopyright{acmlicensed} 
\copyrightyear{2018} 
\acmYear{2018} 
\acmDOI{XXXXXXX.XXXXXXX} 
\acmConference[Conference acronym 'XX]{Make sure to enter the correct
  conference title from your rights confirmation email}{June 03--05,
  2018}{Woodstock, NY}  
\acmISBN{978-1-4503-XXXX-X/2018/06}  

\usepackage{xspace}
\usepackage{longtable}
\usepackage{multirow}
\usepackage{subcaption}
\usepackage{listings}
\usepackage{todonotes}
\usepackage{tcolorbox}
\usepackage{enumitem}
\usepackage{comment}
\usepackage{fancyvrb}
\usepackage{upquote}               
\usepackage{lstlinebgrd}

\usepackage{tikz}
\usetikzlibrary{shapes}
\usetikzlibrary{positioning, fit, backgrounds, calc, bending, tikzmark}

\IfFileExists{fontawesome5.sty}{\usepackage{fontawesome5}}{}

\definecolor{pinkheader}{RGB}{244, 143, 147}
\definecolor{redtext}{RGB}{200, 80, 80}
\definecolor{outerbox}{RGB}{245, 240, 235}
\definecolor{blueheader}{RGB}{135, 175, 230}
\definecolor{bluetext}{RGB}{70, 110, 180}

\definecolor{codebg}{RGB}{250,250,250}
\definecolor{codeframe}{RGB}{220,220,220}
\definecolor{numbers}{gray}{0.50}
\definecolor{diffadd}{RGB}{6,125,34}
\definecolor{diffdel}{RGB}{190,0,0}
\definecolor{diffhunk}{RGB}{0,90,160}
\definecolor{metainfo}{gray}{0.45}

\lstdefinelanguage{diff}{
  morecomment=[f][\color{metainfo}]{diff},
  morecomment=[f][\color{metainfo}]{index},
  morecomment=[f][\color{metainfo}]{---},
  morecomment=[f][\color{metainfo}]{+++},
  morecomment=[f][\color{diffhunk}]{@@},
  morecomment=[l][\color{diffadd}]{+},
  morecomment=[l][\color{diffdel}]{-},
}

\lstdefinestyle{patch}{
  language=diff,
  backgroundcolor=\color{codebg},
  basicstyle=\ttfamily\fontsize{8pt}{10pt}\selectfont,
  columns=fullflexible,
  keepspaces=true,
  showstringspaces=false,
  showtabs=false,
  numbers=left,
  numberstyle=\scriptsize\color{numbers},
  numbersep=8pt,
  frame=single,
  rulecolor=\color{codeframe},
  framerule=0.4pt,
  xleftmargin=2.5em,
  framexleftmargin=1.5em,
  tabsize=2,
  breaklines=true,
  breakatwhitespace=true,
  prebreak=\raisebox{0ex}[0ex][0ex]{\tiny$\hookleftarrow$},
  postbreak=\mbox{\space\tiny$\hookrightarrow$},
  literate=
    *{+}{{{\color{diffadd}+}}}1
    {-}{{{\color{diffdel}-}}}1
     {@@}{{{\color{diffhunk}@@}}}2
}

\lstdefinestyle{codeblock}{
  basicstyle=\ttfamily\footnotesize,
  backgroundcolor=\color{gray!5},
  breaklines=true,
  breakatwhitespace=false,
  columns=fullflexible,
  keepspaces=true,
  showstringspaces=false,
  numberstyle=\scriptsize\color{numbers},
  numbersep=0.4em,
  xleftmargin=1.2em,
  tabsize=2,
  frame=none,
  captionpos=b,
}

\newcommand*\ucircled[1]{\tikz[baseline=(char.base)]{%
  \node[shape=circle,draw,fill=black,text=white,inner sep=0.5pt] (char) {#1};}}

\newcommand{\sys}{\mbox{\textsc{SeedSmith}}\xspace}

\newcommand{\newpar}[1]{\medskip\noindent\textbf{#1.}\xspace}

\begin{document}

\title{\sys: LLM-Driven Seed Synthesis for Directed Fuzzing}

\author{Junmin Zhu*}
\affiliation{%
  \institution{UC Santa Barbara}
  \country{California, USA}}
\email{junmin@ucsb.edu}
\author{Siyu Liu*}
\affiliation{%
  \institution{Arizona State University}
  \country{Arizona, USA}}
\email{sliu274@asu.edu}
\author{Jie Hu}
\affiliation{%
  \institution{Arizona State University}
  \country{Arizona, USA}}
\email{jiehu12@asu.edu}
\author{Fabio Gritti}
\affiliation{%
  \institution{UC Santa Barbara}
  \country{California, USA}}
\email{degrigis@cs.ucsb.edu}
\author{Ati Priya Bajaj}
\affiliation{%
  \institution{Arizona State University}
  \country{Arizona, USA}}
\email{atipriya@asu.edu}
\author{Hulin Wang}
\affiliation{%
  \institution{Arizona State University}
  \country{Arizona, USA}}
\email{hwang551@asu.edu}
\author{Wenbo Guo}
\affiliation{%
  \institution{UC Santa Barbara}
  \country{California, USA}}
\email{henrygwb@ucsb.edu}
\author{Tiffany Bao}
\affiliation{%
  \institution{Arizona State University}
  \country{Arizona, USA}}
\email{tbao@asu.edu}
\author{Christopher Kruegel}
\affiliation{%
  \institution{UC Santa Barbara}
  \country{California, USA}}
\email{chris@cs.ucsb.edu}
\author{Giovanni Vigna}
\affiliation{%
  \institution{UC Santa Barbara}
  \country{California, USA}}
\email{vigna@ucsb.edu}

\renewcommand{\shortauthors}{Anonymous}

\begin{abstract}
Directed fuzzing steers fuzzers toward user-defined sink functions to identify vulnerabilities, but it frequently fails to trigger crashes even after long campaigns. 
We identify two challenges that prevent directed fuzzers from exposing crashes: incomplete static analysis of indirect calls, which leaves reachable paths invisible to distance-based guidance, and lack of semantic guidance for crash preconditions, which blind mutation cannot satisfy within practical time budgets. 
A natural intervention point is the initial seed corpus: seeds that encode the right control-flow path and satisfy key crash preconditions shift fuzzing from blind exploration to local refinement.
Existing seed generation approaches address neither: grammar-based and format-driven methods produce structurally valid inputs with no sink awareness, while LLM-based methods either lack sink targeting or inherit static analysis limitations through one-shot prompting. 
We present \sys, an agentic LLM pipeline that replicates a security analyst's workflow: starting from a sink, \sys iteratively explores the codebase, resolves indirect calls, identifies crash preconditions, and synthesizes concrete inputs that satisfy them. 
Because \sys operates as a seed generation front-end, its seeds are fuzzer-agnostic and improve any downstream mutation-based fuzzer without modification. 
On Magma, fuzzers using \sys seeds achieve geometric mean crash-time speedups of $11.51\times$ (\textsc{AFL++}) to $14.66\times$ (\textsc{AFLGo}) over default seeds. 
On ARVO, \sys enables fuzzers to trigger 16 previously unreachable bugs spanning 10 projects with diverse input formats.
\end{abstract}



\maketitle

\section{Introduction}
\label{intro}

Directed fuzzing~\cite{du2022windranger,huang2022beacon,xiang2024critical,wang_predator_2025,bao_alarms_2025,zhang_predecessor-aware_2024,bohme2017directed} is a powerful technique for uncovering vulnerabilities in specific code regions, with proven applications in static-analysis report verification~\cite{bao_alarms_2025}, crash reproduction~\cite{zhang_predecessor-aware_2024}, and patch testing~\cite{xiang2024critical}.
Unlike coverage-guided fuzzing, which broadly explores a program's state space, directed fuzzers focus their efforts on user-defined \emph{sink functions}, which are locations likely to harbor security-critical flaws.
Yet despite this focus, directed fuzzers frequently fail to trigger vulnerabilities even after long campaigns, even though prior work has significantly advanced their scheduling~\cite{bohme2017directed} and mutation~\cite{wang_predator_2025} strategies.

Two challenges explain why directed fuzzers frequently fail to trigger crashes.
First (\textbf{C1}), many directed fuzzers~\cite{bohme2017directed, huang2022beacon, rong_toward_2024} rely on call graph information from static analysis to compute the distance metrics used to drive the exploration process.
However, indirect calls, pervasive in real-world C/C++ programs through function pointers and virtual dispatch, are notoriously difficult to resolve statically~\cite{sui_svf_2016}.
As a result, when call edges are missing, the fuzzer's distance computation is incomplete, leaving entire reachable paths invisible to the guidance algorithm~\cite{liyanage_reachable_2023}.
Second (\textbf{C2}), even when the fuzzer reaches the sink, triggering the crash often requires satisfying precise input-level conditions, such as specific field values, format constraints, or data-flow dependencies.
These conditions are invisible to coverage-based feedback, and blind byte-level mutation is unlikely to satisfy all constraints simultaneously within practical time budgets~\cite{chenangora2018}.

The natural intervention point for both challenges is the \emph{initial seed corpus}: a seed that already encodes the correct control-flow path and satisfies key crash preconditions reduces the fuzzer's task from blind exploration to local refinement. Yet existing seed generation techniques fall short.
Grammar-based~\cite{blazytko2019grimoire,wang2017skyfire}, format-specification-driven~\cite{dutra2022formatfuzzer}, and general LLM-prompting approaches such as Fuzz4All~\cite{xia2024fuzz4all} produce structurally valid inputs, satisfying the format-validity component of \textbf{C2}, but are sink-agnostic: they do not reason about a specific target, so they they do not address \textbf{C1} and leave sink-specific preconditions unaddressed, such as which particular field values or data-flow patterns trigger the crash.
ISC4DGF~\cite{xu2024isc4dgf} and Magneto~\cite{zhou2024magneto} do target specific sinks, but rely on one-shot prompting with static analysis summaries, directly inheriting their incompleteness and providing no mechanism to iteratively reason about crash preconditions.
What is needed is a seed generation approach that actively explores the codebase, resolves indirect calls that static analysis misses, and reasons about the input conditions required to manifest the bug.

In fact, when trying to reach a vulnerable code location, a skilled security analyst does not rely on a precomputed call graph or a single-shot prompt: starting from a sink function, they iteratively explore the codebase, trace call paths, resolve indirect calls by reading function bodies and type definitions, identify the conditions required to trigger the crash, and synthesize a concrete input that satisfies them. 
Recent advances in LLMs allow us to automate this workflow: the demonstrated ability of LLMs to comprehend code structure, reason about data flows, and understand format specifications~\cite{nam_using_2024, zhang_low-cost_2025} maps directly onto each step of the analyst's process, without the incompleteness that limits static analysis. 

Given this insight, we present \sys, a system that addresses both challenges by using an LLM agent to replicate the workflow of a security analyst performing manual seed construction. 

\sys is designed as a two-stage agentic pipeline.
An \emph{Analysis Agent} reconstructs the execution path from harness to sink under a strict context budget, supported by two design choices that
keep the search tractable.
First, a \emph{path optimization} step
compresses the over-approximated call graph into a single linearized path: prefix and suffix segments shared across alternative routes are preserved verbatim, while their divergent middles are collapsed into a single connector node that the agent can expand on demand. 
The agent thus starts from a compact outline of the reachability structure rather than the full set of paths.
Second, its code-search tool is \emph{context-aware}: rather than returning a fixed-line window around a search result, the tool inspects the match site and returns the enclosing function body, type definition, or configuration block, as appropriate, so each query yields a semantically complete unit rather than a truncated fragment. 
Iterative queries against this tool enable the agent to resolve indirect calls (\textbf{C1}) and extract crash preconditions (\textbf{C2}) that static analysis misses, and to emit a structured report on convergence.

The \emph{Seed Generation Agent} then consumes this report and produces Python scripts that programmatically construct test inputs, using format libraries where raw bytes cannot satisfy structural constraints.
Generated seeds are validated against a sanitizer-instrumented binary in a refinement loop, with execution feedback used to correct them. 
Separating analysis from generation also serves a deliberate cost/capability trade-off: the heavyweight code-reasoning model is invoked only during the exploration (analysis) stage, with a cheaper model handling seed generation.

We evaluate \sys on 23 Magma bugs and 115 ARVO challenges across 26 projects, to our knowledge, the largest evaluation of LLM-based seed generation for fuzzing to date.
On Magma, fuzzers using \sys seeds achieve geometric mean crash-time speedups of $11.51\times$ (\textsc{AFL++}) to $14.66\times$ (\textsc{AFLGo}) over default seeds.
Notably, \textsc{AFL++} with \sys outperforms \textsc{AFLGo} at exposing targeted crashes, showing that \sys can give a general-purpose fuzzer directed-fuzzing capability.
On ARVO, fuzzers using \sys seeds trigger 16 bugs that \textsc{AFLRun} and \textsc{AFL++} with default seeds never trigger, spanning 10 projects with diverse input formats.

\smallskip
\noindent In summary, we make the following contributions:

\begin{itemize}[leftmargin=*]
\item We identify two fundamental challenges, incomplete static analysis of indirect calls (\textbf{C1}) and lack of semantic guidance for crash preconditions (\textbf{C2}), that limit both directed fuzzers and existing seed generation techniques for sink-oriented vulnerability discovery.

\item We design and implement \sys, an agentic LLM pipeline that iteratively explores target codebases to generate structurally valid, sink-targeted seeds, addressing \textbf{C1} and \textbf{C2} without modifying the downstream fuzzer.

\item We conduct a large-scale evaluation on 23 Magma bugs and 115 ARVO challenges across 26 projects, demonstrating significant crash-time speedups and new crash discovery across four fuzzers.
\end{itemize}

\section{Background}
\label{background}



\newpar{Fuzzing}
Fuzzing is a dynamic software testing technique that automatically generates inputs to discover bugs in target programs. 
At the start of a fuzzing campaign, the fuzzer is seeded with at least one initial test case. 
It then repeatedly selects a seed from the corpus, applies mutations to produce new inputs, and executes the target program. 
Inputs that trigger new behavior, such as covering previously unseen code, are retained in the corpus for further mutation.
Directed fuzzing~\cite{bohme2017directed, chen2018hawkeye, du2022windranger} extends this workflow by steering execution toward a predefined set of target locations, known as sinks. 
Rather than maximizing overall coverage, directed fuzzers assign higher priority to seeds whose execution traces are estimated to be closer to the sink, using distance metrics computed over a control-flow graph constructed via static analysis.
This makes directed fuzzing especially effective when prior knowledge has identified specific code locations as likely vulnerability sites, such as recently patched functions or sinks flagged by static-analysis tools.
However, the effectiveness of distance-based guidance depends critically on the completeness of the underlying static analysis: when call edges are missing, distance metrics are unreliable, and the fuzzer cannot steer execution toward paths it cannot see.
Recent work uses LLMs to augment the fuzzer's feedback signal in this setting: Locus~\cite{zhu_locus_2025} employs an LLM agent to synthesize predicates that are instrumented into the target program, steering mutation toward predicate-satisfying inputs without modifying the seed corpus.
\newpar{Seed Generation}
The quality of the initial seed corpus significantly impacts fuzzing effectiveness, especially for programs that accept highly structured inputs such as images, documents, or protocol messages.
Traditional approaches generate seeds using learned grammars~\cite{blazytko2019grimoire}, data-driven probabilistic models~\cite{wang2017skyfire}, or binary format specifications~\cite{dutra2022formatfuzzer}, but require explicit grammar knowledge and do not incorporate information about the target vulnerability.
More recently, LLMs have been applied to seed generation. Fuzz4All~\cite{xia2024fuzz4all} prompts LLMs with language specifications to generate diverse inputs, but is not sink-targeted. ISC4DGF~\cite{xu2024isc4dgf} prompts an LLM in a single shot with a static-analysis summary of the C/C++ target to generate seeds tailored to a directed-fuzzing sink. Magneto~\cite{zhou2024magneto} targets Java dependent-library exploitation: it decomposes a known call chain edge by edge and invokes the LLM once per edge on a statically-sliced focal-code window to produce a JSON seed template.
Both ISC4DGF and Magneto pre-compute static analysis and push the resulting summary or slice into the LLM's prompt, inheriting whatever imprecisions (such as missing
indirect calls and data-flow facts) that analysis produces.
\sys differs by employing an \emph{agentic} LLM pipeline that iteratively queries the codebase through a code-search tool and refines its understanding over multiple steps, letting the LLM resolve indirect calls and infer crash preconditions on demand rather than from a fixed static-analysis snapshot.

\newpar{LLM Agents for Code Reasoning}
Large language models are pre-trained on massive corpora that include billions of lines of code, giving them strong capabilities for code comprehension, data-flow reasoning, and format specification understanding without task-specific fine-tuning~\cite{chen_codex, madaan2023selfrefineiterativerefinementselffeedback}. 
When equipped with tools such as code search or file retrieval, LLMs can operate as agents that iteratively query an external environment, observe results, and refine their reasoning over multiple steps~\cite{yao2023reactsynergizingreasoningacting}. 
This agentic paradigm is well-suited to tasks that require building up context incrementally, such as tracing an execution path through a large codebase, resolving indirect calls by reading type definitions and function bodies, and identifying the precise conditions under which a vulnerability manifests. 
\sys exploits these capabilities to replicate the workflow of a security analyst constructing a crash-triggering input from scratch.

\section{Motivation}
\label{motivation}

We motivate \sys with two real-world bugs that expose the limitations of state-of-the-art directed fuzzers, each isolating one of the challenges from Section~\ref{intro}: An nginx bug for \textbf{C1} and an openjpeg bug for \textbf{C2}. 
\subsection{C1: Indirect Calls Break Static Guidance}

Static-analysis-based distance metrics assume that the call graph is sufficiently complete to measure proximity to the sink. 
In practice, indirect calls through function pointers and virtual dispatch routinely violate this assumption, leaving entire reachable paths invisible to the fuzzer's guidance algorithm.



\begin{figure}[h!]
    \centering
    \begin{subfigure}{\linewidth}
    \centering
    \begin{lstlisting}[style=codeblock, language=diff, numbers=left,
        numbersep=0.4em, xleftmargin=1.2em]
@@ -70,7 +70,7 @@ ngx_sendfile_r(ngx_connection_t *c, ngx_buf_t *file, size_t size)
     lseek(file->file->fd, 0, SEEK_SET);
+    rev = ngx_alloc(NGX_SENDFILE_R_MAXSIZE, c->log);
-    rev = ngx_alloc(size, c->log);
     if ( rev == NULL ) {
         return NGX_ERROR;
    \end{lstlisting}
    \caption{Git diff of the injected bug}
    \label{fig:nginx-diff}
    \end{subfigure}
    \begin{subfigure}{\linewidth}
    \centering
    \begin{lstlisting}[style=codeblock, xleftmargin=1.2em]
GET / HTTP/1.1
Host: localhost
Range: bytes=-r, 0-614
    \end{lstlisting}
    \caption{PoV: the \texttt{-r} flag in the \texttt{Range} header triggers the overflow.}
    \label{fig:nginx-pov}
    \end{subfigure}
    \caption{Motivating example for \textbf{C1}: an \texttt{nginx} bug invisible to static call-graph guidance.}
    \label{fig:nginx}
\end{figure}


 
 


Consider the bug shown in Figure~\ref{fig:nginx-diff}, injected into nginx~\cite{nginx} during the DARPA AIxCC competition~\cite{darpa_aixcc}.
The vulnerability is a heap-buffer overflow in \texttt{ngx\_sendfile\_r}, which allocates a fixed-size buffer using the hard-coded constant \texttt{NGX\_SENDFILE\_R\_MAXSIZE} (Line~3) instead of the actual resource size.
Triggering it requires the \texttt{-r} flag to appear at the correct position in the HTTP Range header (Figure~\ref{fig:nginx-pov}), which routes execution into \texttt{ngx\_sendfile\_r} through a chain of function pointers.

AFLRun~\cite{rong_toward_2024} fails to trigger this bug or even reach \texttt{ngx\_sendfile\_r} after 24 hours. The reason is that \texttt{ngx\_sendfile\_r} is reachable only through indirect calls that static analysis cannot resolve~\cite{sui_svf_2016}: no call edge connects the harness to the sink in the computed call graph, so the distance metric assigns the sink infinite distance and the fuzzer receives no gradient toward it.
No amount of mutation can compensate for a missing call edge; the fuzzer simply cannot see the path.

\begin{figure}[h!]
    \centering
    \begin{subfigure}{\linewidth}
    \centering
    \begin{lstlisting}[style=codeblock, language=C++, numbers=left,
        numbersep=0.4em, xleftmargin=1.2em,
        linebackgroundcolor={\ifnum\value{lstnumber}=9\color{yellow!40}\fi}]
if (p_j2k->m_cp.tw == 1 && p_j2k->m_cp.th == 1 &&
    p_j2k->m_cp.tx0 == 0 && p_j2k->m_cp.ty0 == 0 &&
    p_j2k->m_output_image->x0 == 0 &&
    p_j2k->m_output_image->y0 == 0 &&
    p_j2k->m_output_image->x1 == p_j2k->m_cp.tdx &&
    p_j2k->m_output_image->y1 == p_j2k->m_cp.tdy) {
    /* read tile header; return OPJ_FALSE on failure */
    if (!l_go_on ||
            ! opj_j2k_decode_tile(p_j2k, l_current_tile_no, NULL, 0, p_stream, p_manager)) {
        opj_event_msg(p_manager, EVT_ERROR, "Failed to decode tile 1/1\n");
        return OPJ_FALSE;
    }
    \end{lstlisting}
    \caption{Crash site, gated by six tile-geometry predicates.}
    \label{fig:openjpeg}
    \end{subfigure}
    \begin{subfigure}{\linewidth}
    \centering
    \begin{lstlisting}[style=codeblock, language=Python, numbers=left,
        numbersep=0.4em, xleftmargin=1.2em]
from PIL import Image, ImageDraw
# Create a 200x200 image
img = Image.new('RGB', (200, 200), color='white')
draw = ImageDraw.Draw(img)
# Draw a blue circle
draw.ellipse([50, 50, 150, 150], fill='blue')
# Draw a red rectangle
draw.rectangle([75, 75, 125, 125], fill='red')
# Save as JPG
img.save('poc.txt')
    \end{lstlisting}
    \caption{Python script example emitted by \sys.}
    \label{fig:cpvjpg-python}
    \end{subfigure}
    \caption{Motivating example for \textbf{C2}: an \texttt{openjpeg} crash whose six tile-geometry predicates must all hold.}
    \label{fig:openjpeg-combined}
\end{figure}

\begin{figure*}[htb!]
    \centering
    \includegraphics[width=1\textwidth]{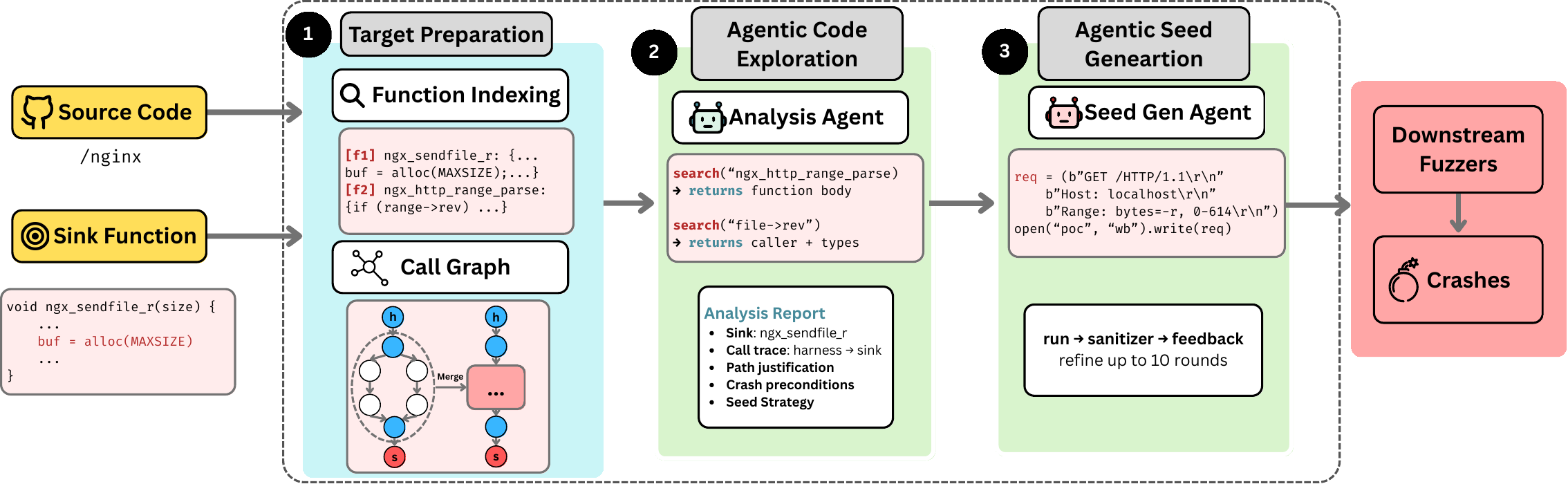}
    \caption{\sys overview. The system processes a C/C++ project through three main stages: \protect\ucircled{1} Target preparation indexes functions, builds call graphs, and instruments the code with sanitizers. \protect\ucircled{2} The Analysis Agent explores the codebase using tool-assisted retrieval to understand how to reach and crash the sink function. \protect\ucircled{3} The Seed Generation Agent creates concrete test inputs based on the analysis, which serve as the initial corpus for any downstream fuzzer.}
    \label{fig:overall}
\end{figure*}

\subsection{C2: Crash Preconditions Defeat Blind Mutation}

Even when the fuzzer successfully reaches a sink, triggering the crash may require satisfying a conjunction of precise input-level conditions that coverage-based feedback cannot distinguish from ordinary reachability.

Consider the bug in openjpeg shown in Figure~\ref{fig:openjpeg}.
The crash site is inside \texttt{opj\_j2k\_decode\_tile} (Line~9), but it is guarded by six tile-geometry predicates (Lines~1--6) that must all hold simultaneously: the tile grid must begin at the origin, the output window must be aligned to the codec's tile size, and the dimensions must match exactly. 
Only when all six conditions are satisfied does execution reach the vulnerable call. 

AFLRun correctly prioritizes inputs that reach the enclosing function, but it never triggers the crash. 
The fuzzer's distance computation provides no signal for satisfying the structural constraints: A seed that reaches the function through the correct path looks identical to one that satisfies all six predicates, from the coverage-feedback perspective. 
Blind byte-level mutation is unlikely to satisfy all constraints simultaneously within practical time budgets~\cite{chenangora2018}, and the vulnerable basic block remains unexecuted throughout the entire 24-hour campaign.


\subsection{Implications for Seed Generation}

These two cases reveal a shared limitation that cannot be overcome by improving the fuzzer's scheduler or mutation strategy alone. When the call graph is incomplete, no distance metric can guide execution toward an unknown path (C1). 
When the crash requires satisfying complex structural predicates, byte-level mutation cannot discover the required input within practical time budgets (C2). 
The natural intervention point is the initial seed corpus: 
A seed that already encodes the correct control-flow path and satisfies key crash preconditions reduces the fuzzer's task from blind exploration to local refinement, making both challenges tractable. 




\section{Design and Overview}
\label{approach}

\sys operates by generating a set of high-quality initial seeds aimed at rapidly triggering potential crashes in targeted code regions. This is achieved by harnessing LLMs' ability to comprehend code structure and data flow, enabling \sys to produce semantically meaningful inputs that sidestep the indirect call resolution limitations inherent in static-analysis-dependent directed fuzzers. 
More importantly, because \sys augments the initial corpus with these seeds rather than replacing any fuzzer component, it is fuzzer-agnostic: the generated seeds can be readily applied to improve both directed fuzzers and coverage-guided fuzzers.
Figure~\ref{fig:overall} presents an overview of \sys's pipeline, which consists of three components described below. 

\ucircled{1} \textbf{Target Preparation}. Taking the target program and sink function as input, this component performs lightweight static analysis and instrumentation to extract program structure and build a queryable knowledge base for the downstream LLM agents.

\ucircled{2} \textbf{Agentic Code Exploration}.
Taking the harness code, sink function, and knowledge base as input, this component employs an LLM agent to explore the codebase and produce a comprehensive analysis report capturing the control-flow paths, data-flow dependencies, and input format requirements needed to reach the target.

\ucircled{3} \textbf{Agentic Seed Generation}. Taking the analysis report as input, this component employs a lightweight LLM agent to generate a set of concrete, semantically meaningful test inputs as the final seed corpus.


The resulting corpus is then handed off to a downstream fuzzer. 
Because \sys operates as a seed generation front-end, it is compatible with any mutation-based fuzzer engine, enabling more directed exploration and faster crash discovery without modifying the fuzzer itself.




\subsection{Target Preparation}

\sys assumes the target is a C/C++ project with available harnesses (e.g., as provided by OSS-Fuzz), and performs three preparation steps: 

{\it 1) Project Indexing}. \sys identifies all functions and their boundaries within the project, building an index that maps function names to their locations and source code for efficient retrieval during downstream code exploration.

{\it 2) Call Path Identification}.
\label{sec:path_opt}
\sys extracts a static call graph from CodeQL.
Direct calls are resolved soundly; indirect calls are resolved by type-signature matching, linking each function-pointer call to every signature-compatible function whose address is taken somewhere in the codebase. This best-effort resolution may both over-approximate, when multiple functions share a parameter signature, and miss edges, when types are obscured by casts or generic pointers. 
\sys deliberately does not require the call graph to be sound: the analysis agent later recovers missing edges through tool-assisted exploration (Section~\ref{agentic_code_exploration}).
From this graph, \sys enumerates candidate paths between the harness and the sink.
Since providing all paths to the LLM would exceed its context limit, \sys applies a \emph{path optimization} step to produce a compact yet informative representation.
Specifically, the algorithm identifies common prefix and suffix segments shared across paths, and merges the divergent middle portions into a single connector node that indicates the existence of multiple branches (illustrated in the \emph{Call Graph} panel of Figure~\ref{fig:overall}).
The result is a linearized path that preserves the essential control flow structure while remaining within the token budget, with the agent retaining the option to explore specific branches on demand via tool-assisted retrieval. 

{\it 3) Project Building}. \sys compiles the project with a sanitizer (\texttt{ASAN}, \texttt{MSAN}, or \texttt{UBSAN}) to enable crash detection during seed validation; the sanitizer is taken from the project's existing build configuration.





\subsection{Agentic Code Exploration}
\label{agentic_code_exploration}

To address the limitations of conventional directed fuzzers, namely incompleteness of static analysis (\textbf{C1}) and semantic information loss (\textbf{C2}), \sys employs an LLM-based \emph{Agentic Code Exploration} component. 
Rather than relying on static analysis alone, an LLM agent iteratively reasons over the identified call paths to understand how the the vulnerability at the sink can be reached and triggered from the harness, ultimately producing a comprehensive analysis report that guides the subsequent agentic seed generation.

Initially, the analysis agent receives three inputs to begin its analysis.

\emph{(I1) Harness code.} The fuzzing entry point of the target project (e.g., from OSS-Fuzz), serving as the starting point for reasoning about execution paths to the sink.

\emph{(I2) Sink function code.} The complete source code of the target function of interest, which may be flagged by a developer, a security patch, or a static analysis tool.



\emph{(I3) Static call path.} The linearized harness-to-sink path produced in step \ucircled{1}. Because the underlying call graph is not sound, \sys treats this as a starting hint rather than ground truth: the path may be (1) empty, if no path exists due to indirect calls or unreachability; (2) a single direct path, if exactly one path is found; or (3) a linearized representation that collapses multiple paths via shared prefix/suffix segments and connector nodes for divergent middles (Section~\ref{sec:path_opt}). 



\newpar{Tool-Assisted Context Retrieval}
The initial inputs (\emph{I1-I3}) rarely suffice to generate valid crash seeds. 
Even when a call path exists, it does not capture the data-flow dependencies and conditional checks that determine whether the sink can be reached and crashed.
Human security researchers address this by iteratively examining the codebase and searching for function definitions, tracing variable assignments, and understanding control-flow conditions. We replicate this process by equipping the agent with a code searching tool.

The search tool accepts keywords or patterns from the LLM and returns contextually enriched results based on the type of match. Rather than returning raw matches with fixed surrounding lines, the tool applies context-aware retrieval rules:
\begin{itemize}[leftmargin=*]
    \item \emph{Function match}: If a match occurs inside a function body, the tool returns the complete function source code, giving the agent full visibility into how the matched symbol is used.
    \item \emph{Global variable match}: The tool returns all global declarations in the file, exposing related state that may affect the execution path.
    \item \emph{Type or structure match}: For matches in source files that correspond to structures, type definitions, or associated comments, the tool returns the relevant declarations with surrounding context.
    \item \emph{Non-source file match}: For configuration files (\texttt{.conf}, \texttt{.ini}, \texttt{.yaml}), the tool returns the entire file; for other non-source files, it returns a localized window around the match.
\end{itemize}
This context-aware retrieval ensures the agent receives semantically meaningful units of code rather than arbitrary line ranges.
For example, when searching for \texttt{file->rev}, if the match appears within \texttt{ngx\_linux\_sendfile()}, the tool returns the entire function's source code to enable analysis of how the variable is used. When searching for a constant like \texttt{NGX\_SENDFILE\_MAXSIZE}, the tool returns surrounding \texttt{\#define} statements that provide context about buffer sizes and related configuration.

Through iterative queries, the agent incrementally accumulates the data- and control-flow dependencies along the execution path.
Each search result may reference new functions, variables, or control-flow conditions, which become targets for subsequent searches. 
This process continues until the agent identifies a complete chain from harness to sink with all necessary preconditions. 
For instance, in \textbf{C1-Example}, the agent identifies the omitted function \texttt{ngx\_http\_range\_parse()}, traces how \texttt{range->rev} is assigned to \texttt{file->rev}, and discovers the critical control-flow condition \texttt{if (file->rev)} that determines whether execution reaches the vulnerable code path. 
The agent also interprets comments and structures to understand input format requirements, discovering that the \texttt{-r} flag in the HTTP Range header can trigger the vulnerability. 
We will show more details about how the analysis agent explores the codebase and finds crashes in Section~\ref{llm_reasoning}.

\newpar{Analysis Report Generation} 
Upon completing its exploration, the agent consolidates its findings into a structured analysis report comprising six components: 
(1) the sink function's location and complete source code; 
(2) the available harness entry points; 
(3) one complete path (call sequence) from the harness to the sink, annotated with the relevant code snippet at each step;
(4) a justification for the selected path over alternatives; 
(5) a detailed analysis of the conditions required to trigger the crash; and (6) a concrete seed generation strategy, including an example Python script. 
This report serves as the sole input to the subsequent Seed Generation Agent (\ucircled{3}), providing all information necessary to produce crash-triggering inputs without further code exploration.


\subsection{Agentic Seed Generation}

The seed generation agent leverages the \emph{Analysis Report} to produce testing inputs (seeds) that attempt to trigger the vulnerability at the sink (\emph{I2}) via the provided harness (\emph{I1}).

Instead of generating raw binary inputs directly, the seed generation agent emits Python scripts that construct the seeds programmatically. 
This design addresses the file structure challenge. Complex file formats such as images, PDFs, or compressed archives require specialized libraries and must satisfy strict structural constraints like headers, checksums, and metadata that LLMs cannot reliably produce as raw bytes. The agent can leverage a set of Python packages, e.g., \texttt{PIL}, to create the target file structure.

Once generated, \sys executes the Python script in an isolated environment to produce raw input seeds. 
To maximize the likelihood of triggering a crash, seed generation follows an iterative refinement loop of up to ten rounds. 
In each round, the generated script is executed, and the resulting seeds are tested against the sanitizer-instrumented target program. 
The loop terminates early if the target crashes on a generated seed; otherwise, the execution feedback (sanitizer output, exit code, and stderr) is passed back to the seed generation agent, which uses it to revise the script for the next round. 
If no crash is detected after ten rounds, all seeds produced across rounds are collected as the final corpus. 
Figure~\ref{fig:cpvjpg-python} shows an example of the script produced.


The resulting seeds are fuzzer-agnostic and can be directly used as an initial corpus for any downstream fuzzer. 
Because the seeds already encode complex reachability conditions such as specific control-flow branches, data-flow dependencies, and input format requirements, they significantly reduce the exploration burden on the fuzzer. 
This benefit holds regardless of the fuzzer's underlying strategy, whether it is directed guidance (e.g., AFLGo~\cite{du2022windranger}), coverage-guided mutation (e.g., AFL++), or otherwise.




\section{Implementation}
\sys was implemented in Python for the core functionality, and it integrates both off-the-shelf and custom tools throughout the various stages of the pipeline.

\smallskip\noindent
\textbf{Target Preparation.}
To prepare the target, we leverage off-the-shelf analysis tools.
In particular, we use \texttt{CodeQL}~\cite{noauthor_codeql_nodate} to build the initial call graph (Section~\ref{sec:path_opt}), and we store such information in a \texttt{Neo4j}~\cite{noauthor_neo4j_2025} graph database to be queried during runtime from the agent at step~\ucircled{1}.
For \emph{Project Indexing}, we implemented a custom C indexer for the extraction of functions and globals.
This indexer is similar to \texttt{tree-sitter}~\cite{tree-sitter}, but extends it with better function boundaries and macro-resolution capabilities.

\smallskip\noindent
\textbf{Agentic Code Exploration.}
We use the \texttt{LangChain} framework~\cite{langchain} to implement the agents in steps~\ucircled{2} and~\ucircled{3}.
Our LLM backends are provided by Anthropic and OpenAI.
We use \texttt{Claude Sonnet 4} for the \emph{Agentic Code Exploration} with default inference parameters (temperature, top-p) because of its state-of-the-art code understanding capability and relatively cheap pricing at the time of evaluation.
We implement the context-aware search tool described in Section~\ref{agentic_code_exploration} using the \emph{grep} tool directly.
While the interface exposed to the LLM is intentionally simple (keyword-based search), the tool internally applies the retrieval rules to enrich raw matches before returning them to the agent.

Specifically, function boundaries and global variable scopes are resolved using the project index built during Target Preparation, and type definitions are extracted from header files with \texttt{X=10} lines of surrounding context for structural matches (\emph{type or structure match}), and \texttt{Y=2} lines for non-source file matches.
These values were chosen empirically: \texttt{X=10} suffices to capture most \texttt{struct}/\texttt{typedef} definitions together with adjacent field comments and related \texttt{\#define}s in typical C/C++ headers, while \texttt{Y=2} provides enough context for key-value entries in configuration files without flooding the agent's context window.

To manage context efficiently, the search tool implements several optimizations.
First, it caches matched lines and enforces deduplication: each unique code snippet appears at most once in the tool output, even if matched multiple times.
Second, if the combined content from all results for a single invocation exceeds 50K tokens, the tool discards the results and returns an informative error message, prompting the agent to issue a more targeted query. 
Returning a truncated subset would risk biasing the agent toward whichever results appear first, whereas an empty result with an explanation preserves the agent's ability to reason about the failure and refine its search.
Third, the agent implements an early termination heuristic: if no matches are found for 30 consecutive tool calls, the analysis halts, and the agent generates its final report.
Since the optimal configuration depends on the specific target under analysis, the values above are the defaults that worked well in our experiments.

\smallskip\noindent
\textbf{Agentic Seed Generation.} 
We use \texttt{gpt-4.1-mini} for the \emph{Agentic Seed Generation}, because of its low-cost but still powerful code generation ability. 
We set the Seed Generation Agent to run ten times.
The generated seeds are validated by executing them against the instrumented binary built during Target Preparation; execution feedback (e.g., unintended crashes or failure to reach the sink) is fed back to the agent for iterative refinement.
%
%
%

\newcommand{\EvalPlaceholder}[1]{\textcolor{blue}{[TBD: #1]}}

\newcommand{\EvalValue}[1]{#1}

\newcommand{\MagmaFastCrashCountAFLpp}{10 } 
\newcommand{\MagmaFastCrashCountFair}{11 } 
\newcommand{\MagmaFastCrashCountAFLRun}{10 } 
\newcommand{\MagmaFastCrashCountAFLGo}{11 } 
\newcommand{\MagmaOneShotCrashCount}{5 } 
\newcommand{\ArvoCrashCountAFLRunCombined}{44 }
\newcommand{\ArvoCrashCountAFLRunDefault}{37 }
\newcommand{\ArvoCrashCountAFLppCombined}{38 }
\newcommand{\ArvoCrashCountAFLppDefault}{26 }
\newcommand{\ArvoCrashGapAFLpp}{12 }
\newcommand{\ArvoCrashGapRateAFLpp}{46\% }
\newcommand{\ArvoCrashGapAFLRun}{4 }
\newcommand{\MagmaCrashGapAFLpp}{3 }
\section{Evaluation}
We conduct a comprehensive evaluation of \sys to demonstrate its effectiveness as an LLM-driven seed generator for directed fuzzing. Our evaluation answers the following research questions:

\begin{enumerate}[label=\textbf{RQ\arabic*:},leftmargin=*]

\item How effective is \sys at generating seeds that crash target sink functions compared to baseline approaches?
\item How does the LLM's reasoning ability help \sys discover crashes faster than state-of-the-art fuzzers?
\item What is the contribution of each component in the \sys design?

\end{enumerate}

We answer \textbf{RQ1} by comparing crash counts and time-to-crash for downstream fuzzers using \sys seeds versus default seeds on ARVO and Magma, and additionally against Locus~\cite{zhu_locus_2025} on Magma (Section~\ref{sec:crash-comparison}).
For \textbf{RQ2}, we manually investigate how LLM reasoning enables \sys to expose crashes significantly faster than baseline fuzzers (Section~\ref{llm_reasoning}).
For \textbf{RQ3}, we present an ablation study quantifying the contributions of the LLM seed generator, the scan strategy, and the control-flow support (Section~\ref{sec:ablation}).
\subsection{Experiment Design}
\label{sec:experiment-design}

\smallskip\noindent
\textbf{Dataset.} We use two benchmarks for a comprehensive evaluation of \sys: Magma and ARVO.

\emph{Magma}~\cite{hazimeh2020magma} is a widely used fuzzing benchmark. 
For our evaluation, we select 23 bugs spanning 8 projects, representing the exact intersection of the datasets used in \textsc{AFLRun} and LibAFLGo~\cite{geretto_libaflgo_2025}. 
This intersection was chosen deliberately: it combines the high vulnerability complexity characteristic of \textsc{AFLRun}'s benchmark with the well-defined, confirmed-reachable target locations rigorously validated by LibAFLGo.
Each bug is instrumented with a unique canary check; a crash is classified as intended when the fuzzer's input triggers that canary.
We run each configuration for 24\,h $\times$ 10 independent trials.

To evaluate real-world applicability, we additionally use \emph{ARVO}~\cite{mei_arvo_2024}, a dataset of over 6,000 reproducible real-world memory vulnerabilities across more than 250 open-source projects, discovered by OSS-Fuzz.
For each bug, ARVO provides a Docker image with a proof-of-concept (PoC) crashing input.
We initially drew a random sample of 200 targets from ARVO to keep the evaluation computationally tractable. 
To ensure a fair comparison and proper execution of directed fuzzing, we filtered out targets that failed to compile under \textsc{AFLRun} or \textsc{AFL++}, yielding a final set of 115 bugs across 26 projects.
A crash is classified as intended when the deduplication token of a fuzzer-generated input matches that of the PoC.
We run each configuration for 24\,h $\times$ 10 trials.

\smallskip\noindent
\textbf{Downstream Fuzzers.}
To demonstrate that \sys seeds are fuzzer-agnostic, we evaluate them with multiple downstream fuzzers.
On Magma we use \textsc{AFL++}(v4.30c), \textsc{AFLGo}~\cite{bohme2017directed}, \textsc{AFLRun}~\cite{rong_toward_2024}, and \textsc{FairFuzz}~\cite{lemieux2018fairfuzz}, covering both directed and coverage-guided strategies.
Although \textsc{AFL++} is designed as a general-purpose fuzzer, prior work has shown that it can outperform existing directed fuzzers at exposing crashes at a target location~\cite{rong_toward_2024,zhang_sok_2026}, a pattern we also observe in our evaluation (see Section~\ref{sec:magma}).
\textsc{AFLGo} is a well-known directed fuzzer commonly used as a baseline, \textsc{FairFuzz} optimizes for rare-branch coverage, and \textsc{AFLRun} is one of the latest directed fuzzers that significantly outperformed eight competitors in its own evaluation.
On ARVO, we use \textsc{AFL++} and \textsc{AFLRun}, selected for their compatibility with OSS-Fuzz projects.

\smallskip\noindent
\textbf{Metrics.}
For each (fuzzer, seed-configuration) pair we measure the \emph{time to trigger} the intended crash, computed as the Restricted Mean Survival Time (RMST)~\cite{hazimeh2020magma} over the 10 trials.
Statistical significance is assessed with one-sided Mann--Whitney $U$ tests; we report $p$-values throughout.
All speedup figures reported in this paper are \emph{geometric means} of per-bug RMST ratios (baseline\,/\,treatment). When only the baseline or only the treatment times out, we substitute 24\,h for its RMST so the ratio remains finite; bugs where both time out are excluded since the ratio is undefined.
In all trigger-time tables, \textbf{bold} marks the fastest configuration per bug per fuzzer (or the faster of the two in the ablation tables), \textbf{T.O.}\ denotes no crash within 24\,h, and \textbf{1-shot} denotes a crash from the LLM-generated seed before any fuzzer mutation (counted as 1\,s when computing speedups).

\smallskip\noindent
\textbf{Environment.}
We conducted experiments on Ubuntu 20.04.6 LTS machines, dedicating one core of an Intel Xeon E5-2670 v2 (2.50 GHz) to each fuzzing campaign.

\smallskip\noindent
\textbf{Seed Configurations.}
For each fuzzer we evaluate four seed configurations:
\emph{No-seed}~(empty initial corpus),
\emph{Default}~(OSS-Fuzz seeds only),
\emph{SS-only}~(\sys-generated seeds only), and
\emph{SS-combined}~(Default\,+\,SS-only).
This design isolates the effect of \sys seeds: improvements under SS-only or SS-combined over Default and No-seed are directly attributable to the quality of \sys-generated seeds.
Unless otherwise noted, ``SS'' and ``\sys seeds'' in the prose refer to \emph{SS-combined}; we use the other configurations only when discussing them explicitly.
\subsection{RQ1: Crash Efficiency}
\label{sec:crash-comparison}
We measure crash counts and time-to-crash with \sys seeds versus Default seeds across all four fuzzers on Magma and across \textsc{AFL++} and \textsc{AFLRun} on ARVO. 
Among prior LLM-augmented directed-fuzzing systems, ISC4DGF~\cite{xu2024isc4dgf} is not open-sourced and Magneto~\cite{zhou2024magneto} targets Java dependent libraries, neither of which can be run on our C/C++ benchmarks. 
We therefore additionally compare against Locus~\cite{zhu_locus_2025} on Magma with \textsc{AFL++} and \textsc{AFLGo}; although Locus intervenes at the fuzzer-feedback layer by instrumenting programs with LLM-synthesized predicates rather than generating seeds, the comparison still isolates the value of \sys's seed-corpus intervention against an alternative LLM-driven approach.


\subsubsection{Results on Magma}
\label{sec:magma}
\begin{table*}[h]
\caption{Crash trigger times on Magma: Default vs.\ SS-only vs.\ SS-combined seeds. $p$-value vs.\ Default in parentheses; \colorbox{green!25}{green} = faster than Default.}
\label{tab:magma_seed_comparison}
\footnotesize
\centering
\setlength\tabcolsep{0pt}
\begin{tabular}{ccccccccccccc}
\toprule
\multirow{2}{*}{Bug ID} & \multicolumn{3}{c}{AFL++} & \multicolumn{3}{c}{FairFuzz} & \multicolumn{3}{c}{AFLGo} & \multicolumn{3}{c}{AFLRun} \\
\cmidrule(lr){2-4}\cmidrule(lr){5-7}\cmidrule(lr){8-10}\cmidrule(lr){11-13}
 & Default & SS-only & SS-combined & Default & SS-only  & SS-combined & Default & SS-only & SS-combined & Default & SS-only & SS-combined \\
\midrule
PDF011 & T.O. & \cellcolor{green!25}\textbf{1-shot}(<.01) & \cellcolor{green!25}\textbf{1-shot}(<.01) & T.O. & \cellcolor{green!25}\textbf{1-shot}(<.01) & \cellcolor{green!25}\textbf{1-shot}(<.01) & T.O. & \cellcolor{green!25}\textbf{1-shot}(<.01) & \cellcolor{green!25}\textbf{1-shot}(<.01) & T.O. & \cellcolor{green!25}\textbf{1-shot}(<.01) & \cellcolor{green!25}\textbf{1-shot}(<.01) \\
PDF018 & T.O. & \cellcolor{green!25}\textbf{1-shot}(<.01) & \cellcolor{green!25}\textbf{1-shot}(<.01) & T.O. & \cellcolor{green!25}\textbf{1-shot}(<.01) & \cellcolor{green!25}\textbf{1-shot}(<.01) & T.O. & \cellcolor{green!25}\textbf{1-shot}(<.01) & \cellcolor{green!25}\textbf{1-shot}(<.01) & 3.25h & \cellcolor{green!25}\textbf{1-shot}(<.01) & \cellcolor{green!25}\textbf{1-shot}(<.01) \\
PDF021 & T.O. & \cellcolor{green!25}\textbf{22.93h}(0.18) & T.O.(1.00) & T.O. & \cellcolor{green!25}\textbf{2.16h}(<.01) & T.O.(1.00) & T.O. & T.O. & T.O. & T.O. & \cellcolor{green!25}\textbf{21.96h}(0.08) & T.O.(1.00) \\
PHP004 & \textbf{12.69h} & T.O.(1.00) & 14.10h(0.69) & 2.33h & T.O.(1.00) & \cellcolor{green!25}\textbf{56.25m}(0.07) & 2.28h & T.O.(1.00) & \cellcolor{green!25}\textbf{74.03m}(0.79) & \textbf{10.57m} & T.O.(1.00) & 13.20m(0.75) \\
PHP009 & 5.44h & \cellcolor{green!25}4.83h(0.03) & \cellcolor{green!25}\textbf{3.17h}(0.52) & T.O. & \cellcolor{green!25}\textbf{17.46h}(0.02) & T.O.(1.00) & 25.54m & \cellcolor{green!25}\textbf{2.86m}(<.01) & \cellcolor{green!25}25.35m(0.38) & 49.25m & \cellcolor{green!25}\textbf{5.20m}(0.07) & \cellcolor{green!25}14.81m(0.16) \\
PNG001 & \textbf{22.78h} & 23.11h(0.34) & T.O.(0.86) & T.O. & T.O. & T.O. & T.O. & \cellcolor{green!25}\textbf{21.79h}(0.18) & T.O.(1.00) & 23.80h & \cellcolor{green!25}22.07h(0.50) & \cellcolor{green!25}\textbf{21.76h}(0.50) \\
PNG007 & 11.51h & 19.09h(0.97) & \cellcolor{green!25}\textbf{4.97h}(0.06) & T.O. & \cellcolor{green!25}\textbf{15.25h}(<.01) & \cellcolor{green!25}21.65h(0.18) & 8.46h & \cellcolor{green!25}7.57h(0.63) & \cellcolor{green!25}\textbf{4.51h}(0.52) & 10.37m & \cellcolor{green!25}10.28m(0.40) & \cellcolor{green!25}\textbf{9.64m}(0.21) \\
SND017 & 12.64h & \cellcolor{green!25}\textbf{13.34m}(<.01) & \cellcolor{green!25}7.58h(0.15) & 22.67h & \cellcolor{green!25}\textbf{24.50s}(<.01) & T.O.(0.94) & 4.91h & \cellcolor{green!25}\textbf{66.32m}(<.01) & \cellcolor{green!25}2.37h(0.21) & 3.21h & \cellcolor{green!25}\textbf{16.07m}(0.01) & \cellcolor{green!25}2.74h(0.65) \\
SND020 & \textbf{58.16m} & 64.59m(0.17) & 74.23m(0.63) & 8.79h & \cellcolor{green!25}\textbf{33.10m}(0.52) & \cellcolor{green!25}8.78h(0.45) & 30.50m & \cellcolor{green!25}15.90m(0.69) & \cellcolor{green!25}\textbf{3.63m}(0.19) & \textbf{1.65m} & 36.33m(0.86) & 2.30m(0.76) \\
SQL002 & \textbf{53.32m} & 2.19h(0.45) & 87.92m(0.86) & \textbf{21.16h} & T.O.(0.94) & T.O.(0.94) & 3.97h & \cellcolor{green!25}17.31m(<.01) & \cellcolor{green!25}\textbf{14.12m}(<.01) & 12.56m & 29.57m(0.15) & \cellcolor{green!25}\textbf{4.16m}(<.01) \\
SQL003 & T.O. & T.O. & T.O. & T.O. & T.O. & T.O. & T.O. & T.O. & T.O. & \textbf{17.32h} & T.O.(1.00) & 22.52h(0.99) \\
SQL012 & T.O. & T.O.(1.00) & \cellcolor{green!25}\textbf{23.23h}(0.18) & T.O. & T.O. & T.O. & \textbf{21.58h} & T.O.(0.94) & T.O.(0.94) & \textbf{9.09h} & 23.61h(1.00) & 11.56h(0.60) \\
SQL013 & T.O. & T.O. & T.O. & T.O. & T.O. & T.O. & T.O. & T.O. & T.O. & T.O. & T.O.(1.00) & \cellcolor{green!25}\textbf{23.25h}(0.18) \\
SQL014 & \textbf{9.96h} & 14.18h(0.79) & 12.44h(0.45) & \textbf{22.78h} & T.O.(0.86) & T.O.(0.86) & 9.70h & \cellcolor{green!25}4.99h(<.01) & \cellcolor{green!25}\textbf{12.72m}(<.01) & 78.36m & \cellcolor{green!25}\textbf{18.48m}(<.01) & \cellcolor{green!25}25.89m(<.01) \\
SQL015 & 23.52h & \cellcolor{green!25}\textbf{19.79h}(0.25) & 23.87h(0.56) & T.O. & T.O. & T.O. & \textbf{20.31h} & 22.41h(0.79) & 21.51h(0.61) & 21.30h & \cellcolor{green!25}16.91h(0.35) & \cellcolor{green!25}\textbf{14.42h}(0.08) \\
SQL020 & 22.13h & \cellcolor{green!25}\textbf{1-shot}(<.01) & \cellcolor{green!25}\textbf{1-shot}(<.01) & T.O. & \cellcolor{green!25}\textbf{1-shot}(<.01) & \cellcolor{green!25}\textbf{1-shot}(<.01) & 21.79h & \cellcolor{green!25}\textbf{1-shot}(<.01) & \cellcolor{green!25}\textbf{1-shot}(<.01) & 4.54h & \cellcolor{green!25}\textbf{1-shot}(<.01) & \cellcolor{green!25}\textbf{1-shot}(<.01) \\
SSL001 & 22.37h & 23.89h(0.95) & \cellcolor{green!25}\textbf{22.29h}(0.91) & T.O. & T.O. & T.O. & \textbf{15.78h} & T.O.(0.99) & 22.02h(0.96) & 3.71h & 16.12h(1.00) & \cellcolor{green!25}\textbf{2.09h}(0.15) \\
SSL020 & 20.33h & T.O.(0.99) & \cellcolor{green!25}\textbf{19.97h}(0.62) & \textbf{19.43h} & T.O.(0.99) & 21.81h(0.87) & T.O. & T.O. & T.O. & \textbf{5.36h} & T.O.(1.00) & 7.96h(0.66) \\
TIF002 & 20.84h & \cellcolor{green!25}\textbf{11.81h}(0.02) & \cellcolor{green!25}11.99h(0.02) & T.O. & \cellcolor{green!25}\textbf{11.59h}(<.01) & \cellcolor{green!25}14.60h(<.01) & 21.00h & \cellcolor{green!25}\textbf{5.50h}(<.01) & \cellcolor{green!25}20.08h(0.33) & 14.32h & \cellcolor{green!25}37.33m(<.01) & \cellcolor{green!25}\textbf{21.69m}(<.01) \\
TIF008 & 22.75h & \cellcolor{green!25}\textbf{3.62h}(<.01) & \cellcolor{green!25}9.18h(<.01) & 21.96h & \cellcolor{green!25}\textbf{5.77h}(<.01) & T.O.(0.86) & T.O. & \cellcolor{green!25}\textbf{9.21h}(<.01) & \cellcolor{green!25}18.82h(0.02) & 17.43h & \cellcolor{green!25}\textbf{2.08m}(<.01) & \cellcolor{green!25}5.27m(<.01) \\
TIF014 & 3.30h & \cellcolor{green!25}\textbf{1-shot}(<.01) & \cellcolor{green!25}\textbf{1-shot}(<.01) & 19.62h & \cellcolor{green!25}\textbf{1-shot}(<.01) & \cellcolor{green!25}\textbf{1-shot}(<.01) & 42.35m & \cellcolor{green!25}\textbf{1-shot}(<.01) & \cellcolor{green!25}\textbf{1-shot}(<.01) & 18.12m & \cellcolor{green!25}\textbf{1-shot}(<.01) & \cellcolor{green!25}\textbf{1-shot}(<.01) \\
XML003 & 8.81h & \cellcolor{green!25}\textbf{1-shot}(<.01) & \cellcolor{green!25}\textbf{1-shot}(<.01) & T.O. & \cellcolor{green!25}\textbf{1-shot}(<.01) & \cellcolor{green!25}\textbf{1-shot}(<.01) & 3.39h & \cellcolor{green!25}\textbf{1-shot}(<.01) & \cellcolor{green!25}\textbf{1-shot}(<.01) & 31.40m & \cellcolor{green!25}\textbf{1-shot}(<.01) & \cellcolor{green!25}\textbf{1-shot}(<.01) \\
XML009 & 11.35h & \cellcolor{green!25}\textbf{6.27h}(0.04) & \cellcolor{green!25}9.67h(0.34) & T.O. & T.O. & T.O. & \textbf{31.21m} & 10.03h(1.00) & 32.40m(0.48) & 3.94h & 6.79h(0.85) & \cellcolor{green!25}\textbf{30.61m}(0.21) \\
\midrule
\textit{Geomean Speedup} & -- & \textbf{12.67$\times$} & \textbf{11.51$\times$} & -- & \textbf{21.01$\times$} & \textbf{12.29$\times$} & -- & \textbf{11.14$\times$} & \textbf{14.66$\times$} & -- & \textbf{7.80$\times$} & \textbf{13.15$\times$} \\
\bottomrule
\end{tabular}%
\end{table*}


\begin{figure}[!htb]
    \centering
        \includegraphics[width=\columnwidth]{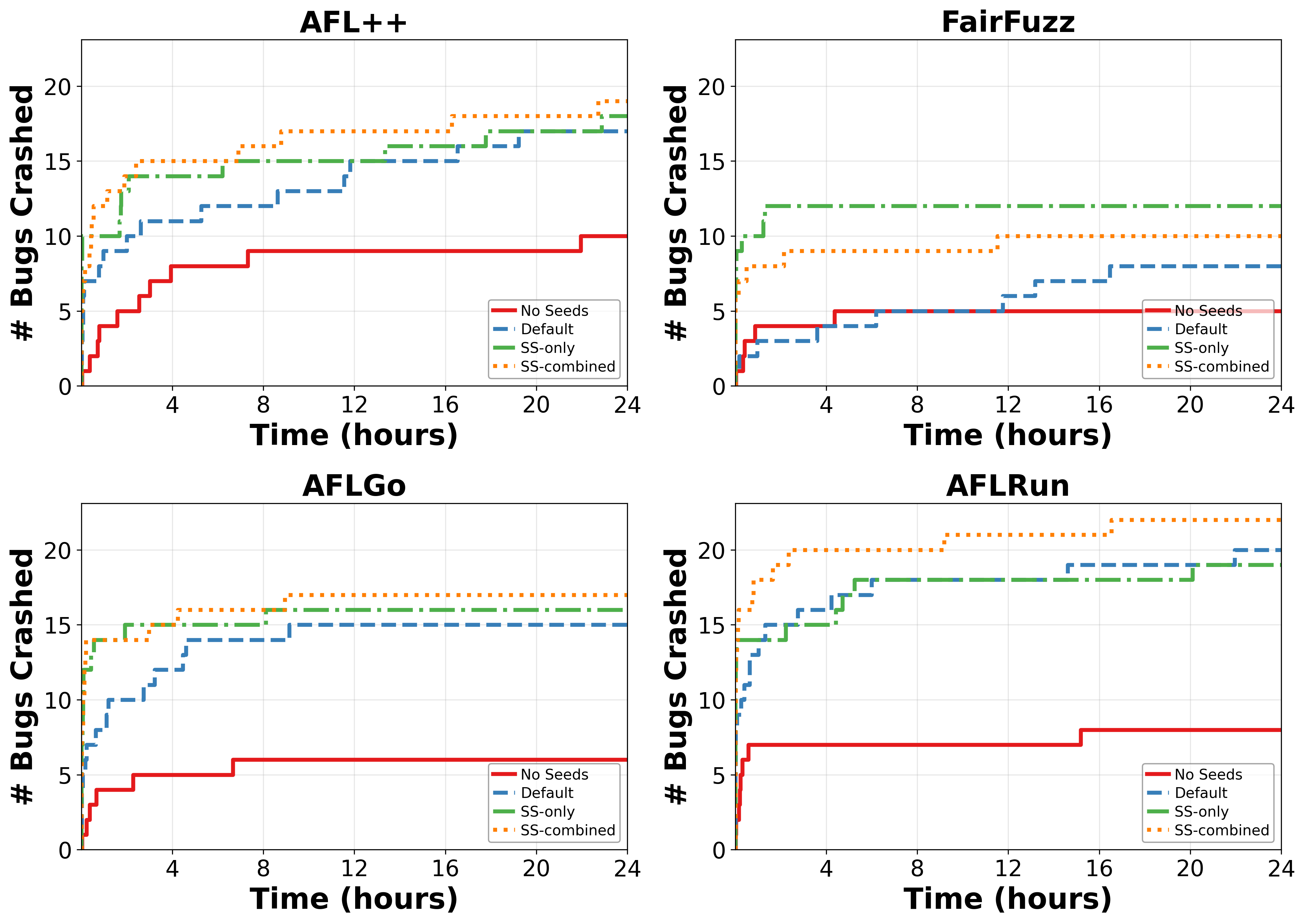}
        
        \caption{Cumulative unique crashes triggered over 24\,h on Magma (23 bugs), one subplot per fuzzer.}
        \label{fig:crashes_magma}
\end{figure}



Table~\ref{tab:magma_seed_comparison} presents trigger times across 23 Magma bugs for all four fuzzers under three seed configurations.
SS-only and SS-combined cells are highlighted \colorbox{green!25}{green} when faster than the corresponding Default cell for the same fuzzer.

With \sys seeds, four fuzzers found 22 out of 23 crashes within 24\,h, versus 20 with Default seeds. The two additional bugs are \texttt{PDF011} and \texttt{SQL013}, neither of which Default ever triggers: \sys crashes \texttt{PDF011} in one shot and triggers \texttt{SQL013} in 23.25\,h with \textsc{AFLRun}. Both reflect \textbf{C2}, since the bottleneck is satisfying the crash predicates at the sink rather than reaching it (Default reaches \texttt{PDF011} in seconds and \texttt{SQL013} in $\sim$16\,h, but neither satisfies the crash condition within 24\,h). Across the 22 \sys-crashed bugs, five (\texttt{PDF011}, \texttt{PDF018}, \texttt{SQL020}, \texttt{TIF014}, \texttt{XML003}) are one-shot crashes where the LLM-generated seed triggers the vulnerability before any fuzzer mutation.

Figure~\ref{fig:crashes_magma} plots cumulative crash counts under all four configurations. 
The No-seed curve consistently lags the others by 5--10 bugs across all four fuzzers, confirming that initial seed quality is a primary determinant of crash discovery on Magma. 
The SS-only and SS-combined curves rise steeply within the first hour and then plateau, whereas Default and No-seed climb gradually throughout the 24\,h budget; this front-loading reflects one-shot and near-one-shot crashes triggered by LLM-generated seeds at $t\approx0$ before any meaningful mutation occurs.
Notably, \textsc{AFL++} with SS-combined achieves the second-best performance among the four fuzzers, suggesting that \sys's targeted seed corpus can improve the directed crash-finding capability of a general-purpose fuzzer. 
This opens the door to a broader range of fuzzers benefiting from \sys seeds, since even non-directed fuzzers can leverage the improved seed quality to trigger deep bugs more efficiently.
And \sys can be used as a drop-in seed generator for existing pipeline without requiring integration into the fuzzer's internals, making it widely applicable across fuzzing ecosystems.

\sys is significantly faster than Default ($p<0.05$) on \MagmaFastCrashCountAFLpp{}of the 23 targets for \textsc{AFL++} and \textsc{AFLRun}, and on \MagmaFastCrashCountFair{}for \textsc{FairFuzz} and \textsc{AFLGo}.
Beyond the \MagmaOneShotCrashCount one-shot cases, the remaining accelerated targets see significant time-to-crash reductions through fuzzer-driven mutation of \sys seeds, indicating that even when \sys does not produce a crashing input directly, the seeds encode useful reachability or crash-predicate hints that give the fuzzer a stronger starting point. 
Geomean speedups under SS-combined range from $11.51\times$ (AFL++) to $14.66\times$ (AFLGo), and under SS-only span $7.80\times$ (\textsc{AFLRun}) to $21.01\times$ (\textsc{FairFuzz}); all four fuzzers show double-digit improvements under at least one configuration.

\begin{table}[h]
    \centering
    \footnotesize
    \setlength{\tabcolsep}{3pt}
    \caption{Crash trigger times on the Locus subset of Magma.}
    \label{tab:locus}
    \begin{tabular}{l rrr rrr}
        \toprule
        & \multicolumn{3}{c}{\textbf{AFL++}} & \multicolumn{3}{c}{\textbf{AFLGo}} \\
        \cmidrule(lr){2-4} \cmidrule(lr){5-7}
        \textbf{Bug} & Locus & Default & SS & Locus & Default & SS \\
        \midrule
        PNG007 & 11.7h & 11.5h & \textbf{5.0h} & 14.1h & 8.5h & \textbf{4.5h} \\
        SND017 & 17.8h & 12.6h & \textbf{7.6h} & 14.1h & 4.9h & \textbf{2.4h} \\
        SQL002 & 5.7h  & \textbf{0.9h} & 1.5h & 1.7h & 4.0h & \textbf{0.2h} \\
        SQL020 & 21.7h & 22.1h & \textbf{1-shot}  & 22.5h & 21.8h & \textbf{1-shot} \\
        SSL020 & 23.6h & 20.3h & \textbf{20.0h} & T.O. & T.O. & T.O. \\
        TIF014 & 12.9h & 3.3h & \textbf{1-shot}  & 22.5h & 0.7h & \textbf{1-shot} \\
        XML003 & 8.3h  & 8.8h & \textbf{1-shot}  & 2.9h & 3.4h & \textbf{1-shot} \\
        XML009 & \textbf{7.9h} & 11.4h & 9.7h & 1.7h & \textbf{0.5h} & 0.5h \\
        \bottomrule
    \end{tabular}

\end{table}

\begin{figure}[!htb]
    \centering
    \includegraphics[width=\columnwidth]{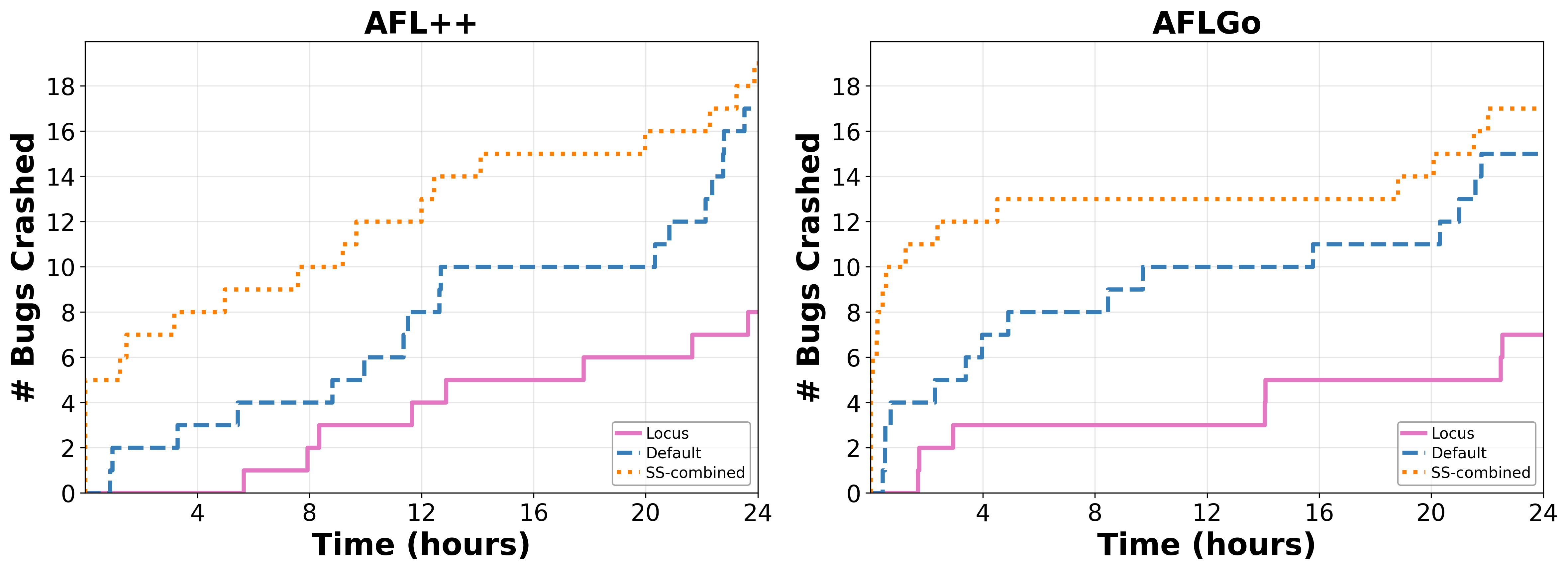}
    \caption{Cumulative unique crashes triggered over 24\,h on Magma, SS-combined vs.\ \textsc{Locus} under \textsc{AFL++} and \textsc{AFLGo}.}
    \label{fig:locus_results}
\end{figure}

A finer-grained pattern emerges when we compare SS-only and SS-combined within each fuzzer family. The two coverage-guided fuzzers achieve their best speedup under SS-only ($12.67\times$ AFL++ and $21.01\times$ FairFuzz), while the two directed fuzzers peak under SS-combined ($14.66\times$ AFLGo and $13.15\times$ AFLRun); the preference is consistent across all four fuzzers.
We attribute this to how each fuzzer family uses its initial corpus: coverage-guided fuzzers rely on seeds for both reachability and exploration, so a focused sink-targeted corpus delivers the largest gain, and adding the default corpus only dilutes the targeting signal.
Directed fuzzers, in contrast, need broad coverage for their distance metric to compute meaningful gradients; SS-only alone exercises too few basic blocks for the metric to be informative, while SS-combined provides both the broad coverage that activates the distance metric and the sink-targeted seeds that lead it directly to the bug. 
\sys therefore narrows the gap between coverage-guided and directed approaches, making the choice of downstream fuzzer less critical to crash-discovery performance, provided the seed configuration is matched to the fuzzer family.

We also compare against Locus~\cite{zhu_locus_2025}, which synthesizes predicates that are compiled into the target rather than generating seeds.
Across all 23 Magma bugs in 24\,hours, \sys triggers 19 crashes on AFL++ and 17 on AFLGo, against Locus's 8 and 7 (Figure~\ref{fig:locus_results}).
On the 8-bug subset where Locus reports results (Table~\ref{tab:locus}), \sys is fastest on 6 of 8 under AFL++ and 6 of 7 valid combinations under AFLGo, with one-shot crashes on \texttt{SQL020}, \texttt{TIF014}, and \texttt{XML003} under both fuzzers.


\subsubsection{Results on ARVO}

\begin{table*}[t]
\caption{Crash trigger times on ARVO: \sys vs.\ Default seeds, only targets crashed at least once are shown.}
\label{tab:arvo_results}
\footnotesize
\centering
\setlength\tabcolsep{1.5pt} %
\begin{tabular*}{\textwidth}{@{\extracolsep{\fill}} l@{\hspace{2pt}}l cccc p{0.1cm} l@{\hspace{2pt}}l cccc p{0.1cm} l@{\hspace{2pt}}l cccc @{}}
\toprule
& & \multicolumn{2}{c}{\textbf{AFL++}} & \multicolumn{2}{c}{\textbf{AFLRun}} & & & & \multicolumn{2}{c}{\textbf{AFL++}} & \multicolumn{2}{c}{\textbf{AFLRun}} & & & & \multicolumn{2}{c}{\textbf{AFL++}} & \multicolumn{2}{c}{\textbf{AFLRun}} \\
\cmidrule{3-4} \cmidrule{5-6} \cmidrule{10-11} \cmidrule{12-13} \cmidrule{17-18} \cmidrule{19-20}
\textbf{Proj.} & \textbf{ID} & Def. & SS & Def. & SS && \textbf{Proj.} & \textbf{ID} & Def. & SS & Def. & SS && \textbf{Proj.} & \textbf{ID} & Def. & SS & Def. & SS \\
\midrule
aom & 42533504 & \textbf{17.14h} & 19.55h & 12.54h & \textbf{10.90h} && libavc & 42531113 & T.O. & T.O. & T.O. & \textbf{23.67h} && libxml2 & 42532747 & T.O. & T.O. & \textbf{21.87h} & T.O. \\
 & 42533540 & \textbf{15.13h} & 16.92h & 12.33h & \textbf{10.29h} &&  & 42531141 & T.O. & T.O. & \textbf{22.42h} & T.O. &&  & 42533922 & \textbf{19.95h} & 20.34h & \textbf{20.79h} & 21.65h \\
assimp & 42528235 & T.O. & \textbf{1-shot} & T.O. & \textbf{1-shot} && libdwarf & 42528680 & 16.25m & \textbf{11.51m} & \textbf{17.30m} & 18.47m &&  & 42533950 & 22.83h & \textbf{22.58h} & 22.05h & \textbf{21.57h} \\
geos & 42532863 & T.O. & \textbf{21.74h} & T.O. & \textbf{23.53h} &&  & 42528792 & 22.23m & \textbf{10.42m} & 50.64m & \textbf{50.54m} &&  & 42534044 & \textbf{17.04h} & 20.37h & T.O. & T.O. \\
inchi & 42534536 & T.O. & \textbf{19.64h} & T.O. & \textbf{6.68h} &&  & 42528883 & 6.17m & \textbf{5.83m} & 32.77m & \textbf{16.21m} &&  & 42534845 & 12.99h & \textbf{7.71h} & 10.16h & \textbf{2.75h} \\
 & 42534760 & T.O. & \textbf{1-shot} & 3.19h & \textbf{1-shot} && libical & 42536107 & \textbf{4.18h} & 16.62h & \textbf{11.45h} & 13.53h &&  & 42537493 & 1.52m & \textbf{49.42s} & 39.27s & \textbf{37.25s} \\
 & 42534959 & T.O. & T.O. & 23.88h & \textbf{22.13h} && libssh2 & 42531314 & T.O. & T.O. & \textbf{5.60h} & 7.62h && openjpeg & 42535258 & T.O. & \textbf{6.32m} & T.O. & T.O. \\
 & 42535235 & T.O. & T.O. & T.O. & \textbf{23.48h} && libtpms & 42537128 & T.O. & \textbf{22.67h} & 17.99h & \textbf{17.51h} && php-src & 42529542 & \textbf{22.12h} & T.O. & T.O. & T.O. \\
 & 42536064 & T.O. & \textbf{1.38m} & 26.97m & \textbf{8.07m} && libxaac & 42527401 & T.O. & \textbf{18.13h} & T.O. & T.O. &&  & 42529650 & T.O. & \textbf{1-shot} & T.O. & \textbf{1-shot} \\
 & 42536079 & T.O. & \textbf{22.64h} & \textbf{14.01h} & T.O. &&  & 42527409 & T.O. & T.O. & \textbf{22.22h} & T.O. && pjsip & 42535147 & T.O. & \textbf{1-shot} & T.O. & \textbf{1-shot} \\
 & 42536250 & T.O. & \textbf{22.28h} & T.O. & T.O. &&  & 42527540 & T.O. & T.O. & \textbf{23.18h} & T.O. && qpdf & 42531940 & T.O. & T.O. & \textbf{22.77h} & T.O. \\
 & 42536593 & T.O. & T.O. & T.O. & \textbf{22.46h} &&  & 42527636 & T.O. & \textbf{23.88h} & T.O. & T.O. &&  & 42535152 & T.O. & \textbf{21.75h} & T.O. & \textbf{21.90h} \\
 & 42536641 & T.O. & \textbf{22.38h} & \textbf{22.46h} & 23.16h &&  & 42531547 & \textbf{1.91h} & 2.82h & \textbf{7.57h} & 18.02h && simdjson & 42534862 & T.O. & T.O. & \textbf{2.40h} & 2.40h \\
lcms & 42529812 & \textbf{19.89h} & T.O. & T.O. & T.O. && libxml2 & 42527008 & 16.76m & \textbf{1-shot} & 28.20m & \textbf{1-shot} &&  & 42534891 & T.O. & T.O. & \textbf{1.00s} & 12.58s \\
 & 42529915 & 1.34h & \textbf{35.28m} & 59.71m & \textbf{40.66s} &&  & 42528997 & \textbf{17.92h} & T.O. & \textbf{15.19h} & 16.19h &&  & 42534894 & T.O. & T.O. & \textbf{1.68s} & 11.43s \\
 & 42529931 & 5.18h & \textbf{2.89h} & 2.67h & \textbf{17.75m} &&  & 42531092 & \textbf{23.26h} & T.O. & \textbf{21.89h} & 23.41h &&  & 42534941 & T.O. & T.O. & \textbf{35.29s} & 2.40h \\
 & 42530151 & 18.86h & \textbf{16.27h} & 15.74h & \textbf{8.43h} &&  & 42531126 & T.O. & \textbf{21.63h} & 21.90h & \textbf{19.54h} && unit & 42536363 & T.O. & \textbf{15.98h} & T.O. & \textbf{1.99h} \\
libavc & 42530446 & T.O. & T.O. & T.O. & \textbf{23.96h} &&  & 42531203 & \textbf{12.01h} & 20.75h & 21.14h & \textbf{15.87h} && xz & 42534913 & 1.16s & \textbf{1-shot} & T.O. & \textbf{1-shot} \\
 & 42530451 & T.O. & T.O. & T.O. & \textbf{22.50h} &&  & 42531481 & \textbf{23.76h} & 23.85h & T.O. & T.O. && zstd & 42532756 & \textbf{20.66h} & 22.06h & 23.82h & \textbf{21.63h} \\
 & 42530453 & \textbf{22.14h} & T.O. & T.O. & T.O. &&  & 42531532 & \textbf{14.47h} & 16.21h & \textbf{17.11h} & 19.76h &&  &  &  &  &  &  \\
\bottomrule
\end{tabular*}
\end{table*}

Table~\ref{tab:arvo_results} reports RMST crash times for the 59 ARVO targets (across 20 projects) where at least one configuration triggered a crash within 24\,h. \sys-combined triggers \ArvoCrashCountAFLRunCombined crashes with \textsc{AFLRun} and \ArvoCrashCountAFLppCombined with \textsc{AFL++}, versus \ArvoCrashCountAFLRunDefault and \ArvoCrashCountAFLppDefault under Default -- a \ArvoCrashGapRateAFLpp{} increase on \textsc{AFL++}, considerably larger than the \MagmaCrashGapAFLpp{}-bug improvement \sys delivers on Magma. We attribute this larger relative gain to ARVO's sparser, less-curated default corpora, which are typical of real-world OSS-Fuzz projects.

\begin{figure}[!htb]
    \centering
        \includegraphics[width=\columnwidth]{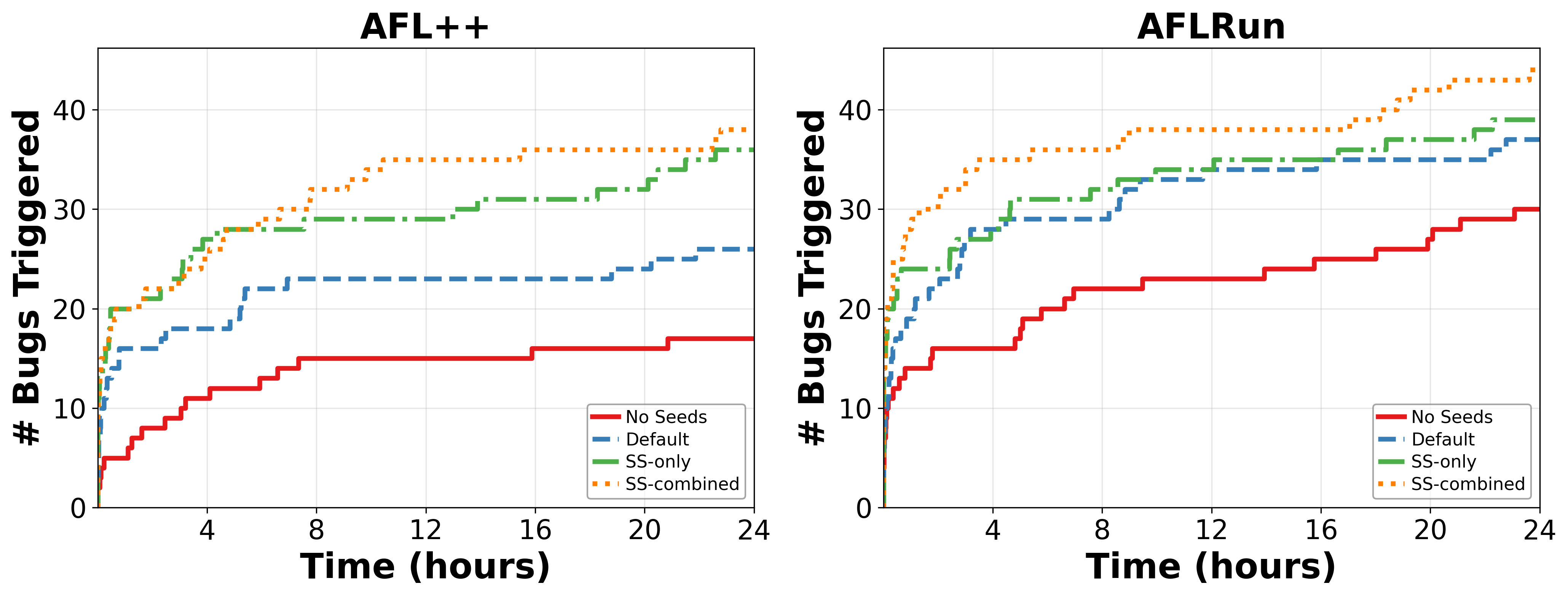}
    \caption{Cumulative unique crashes triggered over 24\,h on 115 ARVO targets, one subplot per fuzzer.}
    \label{fig:crash_arvo}
\end{figure}

Beyond raw counts, \sys unlocks 16 unique crashes across 10 projects that no Default configuration ever triggers: \texttt{assimp}, \texttt{geos}, \texttt{inchi}, \texttt{libavc}, \texttt{libxaac}, \texttt{openjpeg}, \texttt{php-src}, \texttt{pjsip}, \texttt{qpdf}, and \texttt{unit}. The diversity of input formats spanned (binary headers, protocol grammars, compression frames) shows that the LLM's pretraining knowledge of format specifications directly supplies the structural invariants that random mutation cannot construct. Figure~\ref{fig:crash_arvo} shows the same patterns sharpened on ARVO: SS-combined plateaus within 4--6\,h while Default continues climbing, and the layered gap from No-seed to Default to \sys is wider in absolute bugs. 




Across all 59 ARVO bugs (with timeouts capped at 24\,h), \sys seeds yields a $3.09\times$ geomean speedup on \textsc{AFL++} (sign-test $p=0.033$) and a $3.02\times$ speedup on \textsc{AFLRun} ($p=0.043$), both statistically significant at $p<0.05$. This effect is dominated by the 16 unique unlocks where Default never triggers a crash; on the subset of bugs both configurations trigger, the per-target speedup shrinks to $1.46\times$ on \textsc{AFL++} and $1.71\times$ on \textsc{AFLRun} and does not reach significance ($p=0.50$ and $p=0.18$ at $n=21$ and $31$). \sys's primary contribution on ARVO is therefore \emph{unlocking previously unreachable crashes}, with only modest acceleration on shared bugs. This is consistent with the C2 framing, since for already-reachable crashes both configurations face the same predicate-satisfaction problem at the sink.

\begin{table}[!ht]
\centering
\footnotesize
\setlength{\tabcolsep}{3pt}
\caption{Cost and time analysis per project. Counts in parentheses denote the number of targets per project. Rpt. = Analysis report, Seed = Generated seed.}
\begin{tabular}{l  rr  rr  rr}
\toprule
\multirow{2}{*}{Project} & \multicolumn{2}{c}{Total} & \multicolumn{2}{c}{Avg. Rpt.} & \multicolumn{2}{c}{Avg. Seed} \\
& Cost (\$) & Time (s) & Cost (\$) & Time (s) & Cost (\$) & Time (s) \\
\midrule
\texttt{libpng}     (2) & 9.25  & 1373 & 1.47 & 108 & 0.0076 & 13 \\
\texttt{libsndfile} (2) & 9.85  & 1410 & 1.60 & 135 & 0.0046 & 10 \\
\texttt{libtiff}    (3) & 16.38 & 2468 & 1.73 & 131 & 0.0083 & 15 \\
\texttt{libxml2}    (2) & 10.02 & 1161 & 1.62 & 122 & 0.0052 &  7 \\
\texttt{openssl}    (2) & 8.45  & 1132 & 1.35 & 109 & 0.0060 &  8 \\
\texttt{php}        (2) & 8.40  & 1801 & 1.33 & 131 & 0.0074 & 17 \\
\texttt{poppler}    (3) & 10.37 & 1898 & 1.09 & 104 & 0.0067 & 11 \\
\texttt{sqlite3}    (7) & 48.66 & 4357 & 2.28 & 116 & 0.0037 &  9 \\
\midrule
\textbf{Overall} (23) & \textbf{121.38} & \textbf{15600} & \textbf{1.71} & \textbf{118} & \textbf{0.0058} & \textbf{11} \\
\bottomrule
\end{tabular}
\label{tab:cost}
\end{table}

\subsubsection{Seed Generation Cost}
Generating seeds with an LLM incurs additional cost.
Table~\ref{tab:cost} breaks down the cost per project across the 8 Magma projects (per-target measurements are reported in Appendix~\ref{app:cost_per_target}).
The \emph{Agentic Code Exploration} stage takes on average 118\,s and \$1.71 per analysis report when using \texttt{Claude Sonnet 4}, while each seed costs only \$0.0058 and takes 11\,s to generate using \texttt{gpt-4.1-mini}.
On average, a target costs \$5.28 and 678\,s ($\sim$11\,min) of wall-clock time, summing to \$121.38 across all 23 Magma targets.
Compared to a 24\,h fuzzing campaign, this upfront investment is negligible: even the most expensive target (\texttt{SQL012}, \$12.01) costs less than one hour of compute.

\subsection{RQ2: Effectiveness of LLM Reasoning}
\label{llm_reasoning}

We answer \textbf{RQ2} through a manual investigation of \sys's seed-generation process. We use Magma \texttt{PDF018} as a case study (Section~\ref{sec:c1_example}) and then summarize patterns observed across the other accelerated targets (Section~\ref{sec:c1_c2_categorization}).

\begin{figure}[h!]
    \centering
    \begin{lstlisting}[style=codeblock, language=C, numbers=left,
    numbersep=0.5em, xleftmargin=1.5em,
    escapeinside={@}{@},
    linebackgroundcolor={%
      \ifnum\value{lstnumber}=3\color{green!20}\fi
      \ifnum\value{lstnumber}=6\color{green!20}\fi
      \ifnum\value{lstnumber}=13\color{green!20}\fi
      \ifnum\value{lstnumber}=19\color{green!20}\fi}]
// Page.cc -- Page::displaySlice
annotList = getAnnots();
annotList->getAnnot(i)->draw(gfx, printing); // @\textbf{indirect virtual call}@
// Annot.cc -- Annots::createAnnot
} else if (!strcmp(typeName, "Ink")) {
    annot = new AnnotInk(doc, std::move(dictObject), obj);
}
// Annot.cc -- AnnotInk::parseInkList
void AnnotInk::parseInkList(Array *array) {
    for (int i = 0; i < inkListLength; i++) {
        Object obj2 = array->get(i);
        if (obj2.isArray()) {
            inkList[i] = new AnnotPath(obj2.getArray());
    }
}
// Annot.cc -- AnnotInk::draw
for (int i = 0; i < inkListLength; ++i) {
    const AnnotPath *path = inkList[i];       // NULL
    if (path->getCoordsLength() != 0) {       // @\textbf{CRASH: null pointer dereference}@
    }
}
    \end{lstlisting}
    \caption{Simplified \texttt{PDF018} trigger chain: from \texttt{Page::displaySlice} through \texttt{AnnotInk::parseInkList} to the null-pointer dereference in \texttt{AnnotInk::draw}.}
    \label{fig:pdf018-trigger-chain}
\end{figure}

\subsubsection{Case Study}
\label{sec:c1_example}
We use Magma \texttt{PDF018}, a null-pointer dereference in the \texttt{poppler} PDF rendering library, as a case study to illustrate how \sys's LLM-driven analysis addresses both \textbf{C1} and \textbf{C2}.
As Figure~\ref{fig:pdf018-trigger-chain} shows, the crash occurs in \texttt{AnnotInk::draw} (Line~18--19) and is reached through a virtual dispatch from \texttt{Page::displaySlice} (Line~3).

When an annotation has \texttt{/Subtype /Ink}, an \texttt{AnnotInk} object is instantiated (Lines~4--7) and its \texttt{parseInkList} method (Lines~8--15) is called to parse the \texttt{/InkList} array. 
\texttt{parseInkList} zero-initializes the \texttt{AnnotPath*} array via \texttt{memset} and only populates entries whose corresponding PDF element is an array; non-array elements (e.g., \texttt{null}, integers) leave the slot as \texttt{NULL}. \texttt{AnnotInk::draw} (Lines~16--21) then iterates over this array and dereferences \texttt{path->getCoordsLength()} at Line~19 without a null check, resulting in a crash.

This bug exemplifies both challenges. 
For \textbf{C1}, static analysis cannot resolve which concrete \texttt{draw()} override is invoked through the virtual dispatch at Line~3, so directed fuzzers relying on call-graph distance have no guidance toward \texttt{AnnotInk::draw}. 
For \textbf{C2}, constructing the malformed \texttt{/InkList} requires PDF syntax and \texttt{parseInkList} semantics that random mutation cannot discover, but that the LLM draws from its pretraining on format specifications.


\newpar{Reasoning trace}
Figure~\ref{fig:pdf018-trace} traces the agent's reasoning. On \texttt{PDF018}, CodeQL fails to build a database, so the agent receives only \emph{I1} (the harness \texttt{LLVMFuzzerTestOneInput}) and \emph{I2} (the sink \texttt{AnnotInk::draw}); \emph{I3} is empty. 
Reading the sink, the agent observes that \texttt{AnnotInk::draw} iterates over \texttt{inkList} and dereferences \texttt{path->getCoordsLength()} with no null check, and asks when \texttt{inkList[i]} can be null (\emph{Q1}--\emph{Q3}). \emph{Q2}'s retrieval of \texttt{parseInkList} exposes the critical pattern: \texttt{memset(inkList, 0, ...)} zero-initializes the array, then only populates entries where \texttt{obj2.isArray()} holds, so any non-array PDF element silently leaves a null slot. Reading the harness, \emph{Q4}--\emph{Q6} reconstruct the call chain \texttt{LLVMFuzzerTestOneInput} $\to$ \texttt{render\_page} $\to$ \texttt{displayPageSlice} $\to$ \texttt{Page::displaySlice} $\to$ \texttt{AnnotInk::draw}. Combining the two findings, the agent emits a PDF whose \texttt{/InkList} contains a valid coordinate array followed by integer \texttt{42}: \texttt{42} passes the PDF lexer and the array-of-arrays validator (both accept arbitrary objects), but fails \texttt{parseInkList}'s \texttt{isArray} check, leaving \texttt{inkList[1]} null. \texttt{AnnotInk::draw} dereferences it on the first iteration, producing the one-shot crash recorded in Table~\ref{tab:magma_seed_comparison}. 
What this trace surfaces is an interprocedural invariant (zeroed slot $\to$ null deref) revealed only by reading \texttt{parseInkList} and \texttt{draw} together; neither static call-graph distance nor a one-shot prompt over the sink alone would expose it.
\begin{figure}[h!]
\centering
\footnotesize
\begin{tabular}{@{}p{0.95\columnwidth}@{}}
\toprule
\textbf{Inputs.}\quad
\emph{I1} \texttt{LLVMFuzzerTestOneInput} (harness)\quad
\emph{I2} \texttt{AnnotInk::draw} (sink)\quad
\emph{I3} $\emptyset$ \emph{(CodeQL fails on \texttt{poppler})} \\
\midrule
\textbf{Sink-side queries} \emph{(when is \texttt{inkList[i]} null?):} \\
\quad Q1.\ search \texttt{AnnotInk::} \hfill$\to$\ sibling methods (\texttt{parseInkList}, \texttt{draw}) \\
\quad Q2.\ search \texttt{parseInkList} \hfill$\to$\ \texttt{memset(0)}; only array entries populated \\
\quad Q3.\ search \texttt{getCoordsLength} \hfill$\to$\ no null guard before deref \\[2pt]
\textbf{Harness-side queries} \emph{(how is sink reached?):} \\
\quad Q4.\ search \texttt{render\_page} \hfill$\to$\ \texttt{displayPageSlice()} \\
\quad Q5.\ search \texttt{displayPageSlice} \hfill$\to$\ \texttt{Page::displaySlice} \\
\quad Q6.\ search \texttt{Page::displaySlice} \hfill$\to$\ \texttt{annot->draw()} \\
\midrule
\textbf{Synthesized seed.}\quad \texttt{/InkList [[100 150 150 150 200 100], 42]} \\
\quad Integer \texttt{42} fails \texttt{isArray()}, leaving \texttt{inkList[1]} null at draw time. \\
\bottomrule
\end{tabular}
\caption{Reasoning trace for \texttt{PDF018}: starting from raw harness and sink, the agent's six tool-driven queries fan out into sink-side and harness-side sub-goals and converge on a single crash-triggering seed.}
\label{fig:pdf018-trace}
\end{figure}

\smallskip
\noindent
\subsubsection{Analysis of Accelerated Magma Targets}
\label{sec:c1_c2_categorization}
The pattern observed in \texttt{PDF018} generalizes across the other nine targets highlighted in green in Table~\ref{tab:magma_seed_comparison} for \emph{every} fuzzer (\texttt{PDF011}, \texttt{PHP009}, \texttt{PNG007}, \texttt{SND017}, \texttt{SQL020}, \texttt{TIF002}, \texttt{TIF008}, \texttt{TIF014}, \texttt{XML003}). All nine require format-specific structural invariants (\textbf{C2}) that random mutation from generic corpora is difficult to construct: SQLite tagged-union (\texttt{SQL020}), JPEG/EXIF MakerNote vendor-prefix (\texttt{PHP009}), XML external-PE default-option (\texttt{XML003}), PDF xref negative-index (\texttt{PDF011}), WAV \texttt{WAVE\_FORMAT\_EXTENSIBLE} channel-mask (\texttt{SND017}), and so on. Five of the nine (\texttt{TIF002}, \texttt{TIF008}, \texttt{TIF014}, \texttt{XML003}, \texttt{PNG007}) also hide the sink behind function pointers or virtual dispatch (\textbf{C1}): libtiff's codec-dispatch tables, libxml2's SAX callbacks, and libpng's warning/error handlers are populated at runtime and invisible to static call-graph construction. 
Across all nine targets, the analysis agent reads the relevant format and dispatch code together (as in the \texttt{PDF018} trace above) and produces seeds that exercise the correct dispatch chain: directly triggering the crash for the five one-shot cases, and seeding the fuzzer within mutation-reachable distance for the rest.

\subsection{RQ3: Ablation Study}
\label{sec:ablation}

\begin{table}[t]
\centering
\caption{\sys seeds vs.\ \sys seeds w/o Scan Strategy.}
\label{tab:ablation_cmb_scan}
\footnotesize
\setlength\tabcolsep{0pt}
\resizebox{\columnwidth}{!}{%
\begin{tabular}{l cc cc cc cc}
\toprule
 & \multicolumn{2}{c}{\textbf{AFL++}} & \multicolumn{2}{c}{\textbf{FairFuzz}} & \multicolumn{2}{c}{\textbf{AFLGo}} & \multicolumn{2}{c}{\textbf{AFLRun}} \\
\cmidrule(lr){2-3}\cmidrule(lr){4-5}\cmidrule(lr){6-7}\cmidrule(lr){8-9}
Bug & w/o Scan & SS & w/o Scan & SS & w/o Scan & SS & w/o Scan & SS \\
\midrule
PDF011 & T.O. & \textbf{1-shot} & T.O. & \textbf{1-shot} & T.O. & \textbf{1-shot} & T.O. & \textbf{1-shot} \\
PNG001 & T.O. & T.O. & T.O. & T.O. & T.O. & T.O. & 23.4h & \textbf{21.8h} \\
PNG007 & 7.6h & \textbf{5.0h} & T.O. & \textbf{21.6h} & \textbf{4.3h} & 4.5h & \textbf{7.1m} & 9.6m \\
SND017 & \textbf{1-shot} & 7.6h & \textbf{1-shot} & T.O. & \textbf{1-shot} & 2.4h & \textbf{1-shot} & 2.7h \\
SND020 & \textbf{47.3m} & 1.2h & 9.6h & \textbf{8.8h} & 24.0m & \textbf{3.6m} & \textbf{2.1m} & 2.3m \\
SSL001 & \textbf{21.8h} & 22.3h & T.O. & T.O. & T.O. & \textbf{22.0h} & 5.6h & \textbf{2.1h} \\
TIF014 & 13.2m & \textbf{1-shot} & 10.0h & \textbf{1-shot} & 35s & \textbf{1-shot} & 24s & \textbf{1-shot} \\
XML003 & \textbf{1-shot} & \textbf{1-shot} & \textbf{1-shot} & \textbf{1-shot} & \textbf{1-shot} & \textbf{1-shot} & \textbf{1-shot} & \textbf{1-shot} \\
\midrule
\multicolumn{1}{l}{\textit{Geomean Speedup}} & \multicolumn{2}{c}{\textbf{3.0$\times$}} & \multicolumn{2}{c}{\textbf{5.9$\times$}} & \multicolumn{2}{c}{\textbf{3.0$\times$}} & \multicolumn{2}{c}{\textbf{2.1$\times$}} \\
\bottomrule
\end{tabular}
}
\end{table}
\begin{table}[t]
\centering
\caption{\sys seeds vs.\ \sys seeds w/o CodeQL-based control-flow support.}
\label{tab:ablation_cmb_codeql}
\setlength\tabcolsep{0pt}
\resizebox{\columnwidth}{!}{%
\begin{tabular}{l cc cc cc cc}
\toprule
 & \multicolumn{2}{c}{\textbf{AFL++}} & \multicolumn{2}{c}{\textbf{FairFuzz}} & \multicolumn{2}{c}{\textbf{AFLGo}} & \multicolumn{2}{c}{\textbf{AFLRun}} \\
\cmidrule(lr){2-3}\cmidrule(lr){4-5}\cmidrule(lr){6-7}\cmidrule(lr){8-9}
Bug & w/o CodeQL & SS & w/o CodeQL & SS & w/o CodeQL & SS & w/o CodeQL & SS \\
\midrule
PDF011 & T.O. & \textbf{1-shot} & T.O. & \textbf{1-shot} & T.O. & \textbf{1-shot} & T.O. & \textbf{1-shot} \\
PNG001 & T.O. & T.O. & T.O. & T.O. & T.O. & T.O. & 24.0h & \textbf{21.8h} \\
PNG007 & 9.6h & \textbf{5.0h} & T.O. & \textbf{21.6h} & 5.0h & \textbf{4.5h} & 21.1m & \textbf{9.6m} \\
SND017 & 11.9h & \textbf{7.6h} & T.O. & T.O. & 6.4h & \textbf{2.4h} & 3.0h & \textbf{2.7h} \\
SND020 & 1.5h & \textbf{1.2h} & 9.1h & \textbf{8.8h} & 21.6m & \textbf{3.6m} & 3.9m & \textbf{2.3m} \\
SSL001 & \textbf{20.3h} & 22.3h & T.O. & T.O. & \textbf{15.3h} & 22.0h & 9.1h & \textbf{2.1h} \\
TIF014 & 5.4m & \textbf{1-shot} & 1.5h & \textbf{1-shot} & 1.7m & \textbf{1-shot} & 28s & \textbf{1-shot} \\
XML003 & 4.0m & \textbf{1-shot} & T.O. & \textbf{1-shot} & T.O. & \textbf{1-shot} & 27s & \textbf{1-shot} \\
\midrule
\multicolumn{1}{l}{\textit{Geomean Speedup}} & \multicolumn{2}{c}{\textbf{30.2$\times$}} & \multicolumn{2}{c}{\textbf{540.9$\times$}} & \multicolumn{2}{c}{\textbf{71.4$\times$}} & \multicolumn{2}{c}{\textbf{13.8$\times$}} \\
\bottomrule
\end{tabular}
}
\end{table}
To address \textbf{RQ3}, we ablate \sys's three components on Magma. The LLM seed generator's contribution is established in RQ1 against Default seeds; here we isolate two further design choices:
(i)The \emph{scan strategy} is the search tool's context-aware retrieval rules, which return semantically complete units (function body, type declaration, or configuration block) instead of raw grep matches (Section~\ref{agentic_code_exploration}); we ablate it by switching the tool back to raw grep output. 
(ii)The \emph{control-flow support} is the CodeQL-derived, path-optimized harness-to-sink path supplied as input \emph{I3} (Section~\ref{sec:path_opt}); we ablate it by replacing \emph{I3} with an empty path. 
We compare full \sys against both ablated variants on 8 Magma bugs.
CodeQL's database extraction succeeds on only these 8 targets due to build-system incompatibilities common in real-world C/C++ projects~\cite{li_iris_2025}; this is an external limitation of CodeQL, and on the other 15 targets \sys gracefully falls back to the no-CodeQL configuration that RQ1 already evaluates. 
The 8-bug subset still spans six input formats (PDF, PNG, SND, SSL, TIFF, XML), preserving Magma's format diversity. 
The speedups reported below are \emph{marginal} contributions of each component on top of the LLM seed generator's RQ1 baseline, not absolute speedups over Default.
Concretely, each ratio is computed as $T_{\text{w/o component}} / T_{SeedSmith}$ on the same bug-fuzzer pair, isolating how much that component accelerates \sys relative to a version of \sys without it; the speedup of \sys over Default itself is the RQ1 number.

\newpar{Effect of the Scan Strategy}
The scan strategy yields per-fuzzer additional geomean speedups of $2.1\times$ to $5.9\times$ (Table~\ref{tab:ablation_cmb_scan}).
Its value is most apparent on targets with multiple plausible execution paths, where directing the agent toward the right one avoids exploratory overhead.
On \texttt{TIF014}, \sys achieves one shot for all four fuzzers while \sys w/o Scan Strategy requires 24\,s--13.2\,m across the directed/coverage fuzzers and up to 10.0\,h for \textsc{FairFuzz}; on \texttt{PDF011}, \texttt{PNG001}, and \texttt{SSL001}, \sys either uniquely triggers the crash or matches the ablated variant.
\texttt{SND017} is the principal exception: \sys w/o Scan Strategy achieves one-shot across all fuzzers while \sys requires 2.4\,h--7.6\,h, because the vulnerability is directly reachable through a single call chain and the scan strategy's path exploration introduces unnecessary indirection.

\newpar{Effect of Control Flow Support}
Having an initial call graph to refine rather than reconstruct from scratch substantially reduces the exploration burden.
The component yields per-fuzzer additional geomean speedups of $30.2\times$ on \textsc{AFL++}, $71.4\times$ on \textsc{AFLGo}, and $13.8\times$ on \textsc{AFLRun} (Table~\ref{tab:ablation_cmb_codeql}), driven primarily by \texttt{PDF011} and \texttt{XML003}, where the ablated variant times out while \sys crashes in one shot.
The spread reflects how much each fuzzer relies on a complete static call graph: \textsc{AFLGo}'s distance metric depends entirely on it, so missing indirect edges blinds its scheduling, while \textsc{AFLRun} augments the call graph at runtime as new edges are observed, partially compensating for the missing edges when CodeQL is unavailable.
\textsc{FairFuzz} is reported separately at $540.9\times$, but this number is not directly comparable: without control-flow support \textsc{FairFuzz} fails to trigger most bugs within 24\,h, so only \texttt{TIF014} and \texttt{XML003} contribute non-T.O.\ ratios, dominating the geometric mean.
The main exception is \texttt{SSL001}, on which the ablated variant is faster for \textsc{AFL++} and \textsc{AFLGo}; the SSL vulnerability hinges on a protocol state-machine error driven by input semantics, so extra call-graph information does not help.

\newpar{Summary}
The two components contribute incrementally and non-redundantly: the scan strategy improves the quality of information the agent receives per code-search query, while control-flow support gives the agent an initial call graph to refine rather than reconstruct from scratch. Both yield positive marginal speedups across all four fuzzers (with the SND017/SSL001 exceptions noted above), and their combination produces the strongest \sys variant evaluated in RQ1.

\section{Discussion and Limitations}
\label{discussion}

\newpar{LLM Seed Generation}
\sys's ability to generate crashing test cases highly depends on the LLM's code-understanding ability.
When LLMs fail to correctly identify the root cause of the vulnerability in the sink function or misdiagnose the trigger condition, the probability of generating a crashing input drops.
This can be mitigated by fine-tuning a model with enhanced vulnerability-analysis capability.
This dependency is also an advantage: as LLMs continue to improve at code understanding and vulnerability reasoning, \sys's seed-generation quality improves correspondingly without any changes to the system itself.

\newpar{Precision of the Sink Function}
In our evaluation, we provide the first function on the crashing stack trace to \sys as the sink function.
However, the root cause of the vulnerability might not be in the provided function.
The real vulnerability might be located in the other functions of the stack trace, or it might not be on the stack trace at all.
By merely targeting the provided sink function without providing any context or information about the actual vulnerability, the LLM is less ineffective.
For \sys, this can be mitigated by providing additional information to the LLM.
For example, for crash reproduction, giving to the LLM the original crash report can make the LLM aware of the possibility that the vulnerability might exist in other functions on the stack trace, thus rendering it easier for LLM to locate the root cause of the vulnerability by potentially exploring more functions on the stack trace.

\newpar{Zero-Day Discovery}
Our evaluation targets the N-day reproduction setting, where the sink function is given as input to \sys.
\sys does not perform sink localization on its own, so applying it to zero-day discovery requires pairing it with a sink-discovery component, such as static bug-pattern detection~\cite{haller2013dowsing} or learned vulnerability symptoms~\cite{Meng2021_Bran}. 
We have evaluated \sys as a separate component in our AIxCC finals submission, paired with a customized setup for confirming the sink function. 
In that deployment, \sys seeds independently one-shot triggered 20 different targets, and further helped downstream fuzzers reach seven zero-day vulnerabilities, four of which were not discovered by any other team.
This demonstrates that \sys generalizes to zero-day discovery when paired with a sink-discovery component.

\section{Related Work}
\label{related_work}
\newpar{Directed Fuzzing}
Directed fuzzing aims to improve the efficiency of vulnerability discovery by steering the fuzzer toward specific, potentially vulnerable code regions. 
These target regions may include code that performs memory operations~\cite{haller2013dowsing}, calls to known vulnerable functions~\cite{wang2025predator}, code matching learned bug symptoms~\cite{Meng2021_Bran}, recently patched code~\cite{marinescu2013katch}, or newly committed changes~\cite{xiang2024critical}. By narrowing the focus, directed fuzzing enhances the likelihood of uncovering security-critical bugs in a more targeted and resource-efficient manner.

Despite variations in how target locations are determined~\cite{osterlund2020parmesan,marinescu2013katch,haller2013dowsing,xiang2024critical,wang2025predator}, directed fuzzers generally share the goal of prioritizing program paths that are semantically or syntactically closer to the targets. Prior work in this area can be broadly categorized into three groups based on the stage of the fuzzing pipeline they aim to optimize: seed scheduling, seed mutation, and mutation tracing.

Among these, seed scheduling has received significant attention. AFLGo~\cite{bohme2017directed} pioneered the approach of assessing seeds based on the distance between their execution traces and the target locations at the basic block level, prioritizing seeds with shorter distances for mutation. Subsequent works refined this distance metric using more precise basic block–level granularity~\cite{du2022windranger,kim2023dafl}, function-level abstractions~\cite{chen2018hawkeye}, data  distances~\cite{lee2021constraint}, and even inter-target correlations~\cite{huang2024titan,zheng2023fishfuzz,liang2023multiple}, all aiming to guide the fuzzing process more effectively toward the target locations.

Seed mutation strategies have also been explored to enhance the likelihood that generated inputs reach target code regions. FairFuzz~\cite{lemieux2018fairfuzz} introduces a mutation mask that biases mutations toward input bytes associated with rare branches, thereby increasing the probability of exercising hard-to-reach paths. RDFuzz~\cite{ye2020rdfuzz} further improves this by identifying and preserving input content that is sensitive to distance metrics, generating mutations more likely to minimize the distance to targets. In parallel, symbolic execution tools~\cite{yun2018qsym,poeplau2020symbolic,cadar2008klee,godefroid2005dart} can generate inputs that precisely satisfy the constraints of a path leading to a target. However, symbolic execution suffers from poor scalability when the target is deeply embedded in the program state space, as it must first enumerate or search a vast number of paths to isolate one that reaches the goal~\cite{ma2011directed,baldoni2018survey}.

Finally, prior work on optimizing the mutation tracing stage focuses on evaluating whether generated inputs are likely to reach the target and pruning those that are not. 
FuzzGuard~\cite{zong2020fuzzguard} trains a neural network to predict target reachability and filter out unpromising inputs before execution. Beacon~\cite{huang2022beacon} uses lightweight static analysis to infer input preconditions and discard infeasible mutations. 

\newpar{LLM-Augmented Fuzzing}
Recent advances in LLMs show that they can synthesize high-quality structured text, which can be used to enhance fuzzing performance from multiple aspects.
The most straightforward usage of LLMs is seed generation. Several prior works~\cite{deng_large_2023, deng_large_2024, xia2024fuzz4all, eom_fuzzing_2024} leveraged the LLM's superior code generation capability to produce fuzzing inputs for programming language compilers, interpreters, and library APIs.
LLMs have also been used to directly generate commands or inputs to popular applications as an initial seed corpus~\cite{asmita_fuzzing_2024}.
Magneto~\cite{zhou2024magneto} took a step further by feeding the LLM fine-grained program structural information retrieved through static analysis to help it generate initial inputs for directed fuzzing of Java dependency libraries.

Although LLMs excel at generating textual inputs, they are less effective when targets consume non-textual inputs.
Instead of directly using the LLM's outputs as fuzzing seeds, G2FUZZ~\cite{zhang_low-cost_2025} instructs the LLM to emit Python scripts that act as input generators and mutators; ProphetFuzz~\cite{wang_prophetfuzz_2024} drives an LLM-based configuration fuzzer that explores option combinations. 
and \textsc{ChatAFL}~\cite{meng_large_2024} uses an LLM to extract protocol grammar from RFC-style documentation.
Orthogonal to fuzzing input generation, LLMs also excel in fuzzer driver generation~\cite{liu_oss-fuzz-gen_2024, zhang_how_2024, lyu_prompt_2024}.
LLMs are pre-trained on open-source projects, which makes them good at generating fuzzer drivers that can explore uncovered library code.




\section{Conclusion}
In this paper, we present \sys, a system designed to enhance vulnerability discovery in code regions surrounding sink functions. 
We demonstrate that \sys effectively mitigates the ``cold-start'' problem of modern fuzzers, significantly reducing both time-to-reach and time-to-crash.


We evaluated \sys on 23 Magma bugs and 115 ARVO challenges across 26 projects.
In the Magma benchmarks, the generated seeds allowed fuzzers to identify 22 of the 23 bugs. This outperformed default seeds, which triggered 20 bugs. Additionally, the system delivered significant geometric mean speedups, specifically $11.51\times$ for AFL++ and $14.66\times$ for AFLGo.
On ARVO, fuzzers using \sys seeds trigger 16 bugs that \textsc{AFLRun} and \textsc{AFL++} with default seeds never trigger, spanning 10 projects with diverse input formats.
While the time-to-crash speedup on bugs that both configurations trigger is not statistically significant ($p=0.58$), the primary advantage of \sys lies in expanding the set of reachable crashes, enabling fuzzers to trigger vulnerabilities that mutation alone cannot reach within the time budget.

\label{conclusion}


\bibliographystyle{ACM-Reference-Format}
\bibliography{references}

\appendix 
\section{Ethics Considerations}

In this work, we present \sys, a tool that can be used for vulnerability discovery. As our work operates on the ARVO dataset, a curated benchmark of known vulnerabilities, no new bugs have been uncovered and, thus, no vulnerability disclosure was necessary.
\section{Generative AI usage}


In our work, LLMs were used for editorial purposes in this manuscript, and the authors inspected all outputs to ensure accuracy and originality.

We used \textsc{Claude Code} to generate utility scripts to run fuzzing campaigns for baselines and processing raw experiment data.
All the running results and experiment data were manually inspected by humans to make sure they are correct.

\section{Per-Target Cost and Time Breakdown}
\label{app:cost_per_target}

Table~\ref{tab:cost_per_target_breakdown}  reports the per-target cost and time
measurements that back the per-project aggregation in Table~\ref{tab:cost}.
Each row corresponds to a single Magma target. \emph{Total} is the sum of
cost/time over all reports for that target; \emph{Avg. Rpt.} is the average
cost/time per analysis report; \emph{Avg. Seed} is the average cost/time
per generated seed.

\begin{table}[H]
\centering
\footnotesize
\setlength{\tabcolsep}{3pt}
\begin{tabular}{l  rr  rr  rr}
\toprule
\multirow{2}{*}{Target} & \multicolumn{2}{c}{Total} & \multicolumn{2}{c}{Avg. Rpt.} & \multicolumn{2}{c}{Avg. Seed} \\
& Cost (\$) & Time (s) & Cost (\$) & Time (s) & Cost (\$) & Time (s) \\
\midrule
PDF011 & 2.11 & 584 & 0.65 & 89 & 0.0056 & 11 \\
PDF018 & 4.50 & 622 & 1.43 & 118 & 0.0070 & 9 \\
PDF021 & 3.76 & 692 & 1.18 & 105 & 0.0074 & 13 \\
PHP004 & 4.19 & 912 & 1.32 & 139 & 0.0077 & 17 \\
PHP009 & 4.21 & 889 & 1.33 & 122 & 0.0071 & 17 \\
PNG001 & 3.78 & 643 & 1.19 & 99 & 0.0073 & 12 \\
PNG007 & 5.47 & 730 & 1.74 & 117 & 0.0079 & 13 \\
SND017 & 5.90 & 726 & 1.92 & 142 & 0.0045 & 10 \\
SND020 & 3.95 & 684 & 1.27 & 128 & 0.0047 & 10 \\
SQL002 & 6.71 & 584 & 2.21 & 131 & 0.0031 & 6 \\
SQL003 & 7.01 & 1121 & 2.29 & 109 & 0.0051 & 26 \\
SQL012 & 12.01 & 503 & 3.98 & 131 & 0.0022 & 4 \\
SQL013 & 4.60 & 531 & 1.48 & 95 & 0.0052 & 8 \\
SQL014 & 8.69 & 563 & 2.86 & 114 & 0.0032 & 7 \\
SQL015 & 5.59 & 544 & 1.83 & 124 & 0.0036 & 6 \\
SQL020 & 4.05 & 511 & 1.31 & 107 & 0.0035 & 6 \\
SSL001 & 4.67 & 530 & 1.50 & 115 & 0.0060 & 6 \\
SSL020 & 3.78 & 602 & 1.20 & 103 & 0.0060 & 10 \\
TIF002 & 5.55 & 723 & 1.77 & 124 & 0.0082 & 12 \\
TIF008 & 6.07 & 723 & 1.95 & 114 & 0.0068 & 13 \\
TIF014 & 4.76 & 1022 & 1.48 & 155 & 0.0100 & 19 \\
XML003 & 4.53 & 530 & 1.46 & 114 & 0.0049 & 6 \\
XML009 & 5.49 & 631 & 1.78 & 130 & 0.0055 & 8 \\
\midrule
\textbf{Avg.} & \textbf{5.28} & \textbf{678} & \textbf{1.71} & \textbf{118} & \textbf{0.0058} & \textbf{11} \\
\bottomrule
\end{tabular}
\caption{Cost and time analysis per target (Cost in USD, Time in seconds).}
\label{tab:cost_per_target_breakdown}
\end{table}

\section{Sample Analysis Report for PDF018}
\label{app:pdf018_report}

This appendix reproduces the  analysis report \sys's agent produced for the \texttt{PDF018} case study (Section~\ref{sec:c1_example}), with reformatting for readability. Boilerplate fields that overlap with the main paper (sink source, harness entry point, full call trace) are omitted; we keep the agent's reasoning steps, prior-round feedback, condition analysis, seed-construction strategy, and the Python script that produced the one-shot crashing seed.

\subsection*{Reasoning for Path Choice}
\begin{itemize}[leftmargin=*, itemsep=2pt]
\item \textbf{Step 1.} Searched \texttt{AnnotInk::} to understand the class structure; found the constructor, \texttt{parseInkList}, and \texttt{draw} methods, revealing that \texttt{inkList} is an array of \texttt{AnnotPath} pointers that can contain null entries.
\item \textbf{Step 2.} Searched \texttt{parseInkList}; the function body exposed the critical pattern:
\begin{lstlisting}[style=codeblock, language=C++, xleftmargin=1em]
void AnnotInk::parseInkList(Array *array) {
    inkListLength = array->getLength();
    inkList = (AnnotPath **)gmallocn(inkListLength, sizeof(AnnotPath *));
    memset(inkList, 0, inkListLength * sizeof(AnnotPath *));
    for (int i = 0; i < inkListLength; i++) {
        Object obj2 = array->get(i);
        if (obj2.isArray())            // only array entries populated
            inkList[i] = new AnnotPath(obj2.getArray());
        // non-array entries leave inkList[i] = null
    }
}
\end{lstlisting}
\item \textbf{Step 3.} Searched \texttt{getCoordsLength}; \texttt{AnnotPath::getX/getY} have bounds checks, but the issue is dereferencing a null \texttt{path} pointer \emph{before} those methods are called.
\item \textbf{Step 4.} Searched \texttt{render\_page} and \texttt{displayPageSlice} to trace the call path from harness to sink, confirming the standard PDF rendering pipeline.
\item \textbf{Step 5.} Searched \texttt{Page::displaySlice} to locate the exact \texttt{annot->draw()} call site.
\end{itemize}

\subsection*{Conditions Analysis}
\textbf{Critical crash point:} \texttt{const AnnotPath *path = inkList[i]; if (path->getCoordsLength() != 0)} dereferences \texttt{path} with no null check.

\smallskip\noindent\textbf{Root cause.} \texttt{parseInkList} (a) allocates \texttt{inkList} and zero-initializes via \texttt{memset}; (b) populates only entries where \texttt{obj2.isArray()} holds; (c) leaves slots for non-array entries (null, numbers, strings, etc.) at zero; (d) \texttt{AnnotInk::draw} subsequently iterates and dereferences \texttt{inkList[i]} without a null check. The triggering conditions are: a null \texttt{inkList[i]} from a non-array \texttt{/InkList} element, the direct dereference in \texttt{draw}, and an \texttt{inkListLength} that includes the null slot.

\subsection*{Seed-Generation Strategy}
The crash is triggered by a PDF with an \texttt{Ink} annotation whose \texttt{/InkList} mixes a valid coordinate array with a non-array entry (\texttt{[valid\_array, 123]} or \texttt{[valid\_array, "s"]}, etc.). \texttt{parseInkList} creates an \texttt{AnnotPath} for the first entry and leaves the second null; \texttt{draw} then crashes at \texttt{path->getCoordsLength()} on the second iteration.




\section{LLM Prompts Example}
\label{appendix:prompt_example}
System prompt of Code Exploration Agent is shown in Figure~\ref{fig:code_agent_prompt}.
\begin{figure*}[t]
\begin{tcolorbox}[
    colback=white,
    colframe=black!30,
    arc=5mm,
    boxrule=1pt,
    left=5pt,
    right=5pt,
    top=5pt,
    bottom=5pt
]

\begin{tcolorbox}[
    colback=pinkheader,
    colframe=pinkheader,
    arc=1mm,
    boxrule=0pt,
    left=4pt,
    right=4pt,
    top=3pt,
    bottom=3pt
]
    {\bfseries\color{white} System Prompt (Code Exploration)}
\end{tcolorbox}
\vspace{4pt}
\begin{Verbatim}[fontsize=\scriptsize, 
                 baselinestretch=0.85, commandchars=\\\{\}, ]
You are a security expert specializing in vulnerability research. Your task is to generate a detailed and precise report on 
how to trigger crashes in sink functions within a target codebase. Your report will be used by a downstream team with NO 
PRIOR KNOWLEDGE OF THE TARGET OR THE CRASH. Therefore, it must include ALL 
NECESSARY INFORMATION:
1. Code snippets
2. Step-by-step explanations
3. Analysis of control-flow paths, relevant conditions, and seed generation strategies
4. Clear instructions for reproducing the crash
\textbf{## Essential Guidelines}:
Tool Usage:
+ Use the provided grep tool (grep -rnE <expression>) to thoroughly analyze the entire source code directory.
+ Employ generalized, yet precise patterns. For instance, use patterns like ->param instead of specific references like 
  obj->param to ensure broader and more inclusive search coverage.
+ Avoid reusing patterns from previous searches.
\textbf{## Report Structure}:
Your report must be structured to include the following sections:
1. Sink function details: location, signature, purpose
2. Harnesses: file names, line numbers, entry points
3. Call trace: from harness to sink (if provided), or discovered via analysis
4. Relevant conditions: if, switch, etc. on the path
5. Seed generation strategy: Use libraries (base64, zlib, etc.) where applicable, and Python code (or pseudocode) for 
   long/complex seeds. The size of the seed MUST be under 2MB.
6. IMPORTANT: THIS GENERATED REPORT MUST NOT CONTAIN REPETITIVE PATTERNS TO DEMONSTRATE A SEED THAT COULD EXPLOIT THE 
   VULNERABILITY. IF YOU HAVE TO SHOW SUCH EXAMPLES, PLEASE PROVIDE A SHORTER SUMMARY.

After completing the analysis, you MUST output a report in the specified format. Your final report must be 
meticulously structured, explicit, and thorough to enable seamless seed generation and crash reproduction by 
the downstream team.
<report>
    <sink_function_details>
        - Location: file path
        - Signature: function declaration
        - Purpose: functionality description
        - Code: complete code snippet of the sink function
    </sink_function_details>
    <harnesses>
        - File Names:
        - [file_name]:[line_number] (entry point)
        - Clearly describe each harness entry point.
        - These are potential harnesses that can be used to trigger the sink function. You can only use one harness per time.
    </harnesses>
    <call_trace>
        - Step-by-step path from harness to sink function
        - Mention explicitly any optimized or omitted intermediate functions
        - MUST Include complete code snippets for each step in the call trace, DO NOT OMIT ANY CODE NECESSARY
        - Use analysis tools as needed to clarify uncertain paths
        - Do not use __connector__ here, use a real path instead
    </call_trace>
    <reasoning_for_path_choice>
    - Justify the selected call trace path over alternatives.
    - Explain step-by-step how you determined this path as the crash path using tool calls:
        - Step 1: patterns to grep, useful information in grep results, why you choose this pattern
        - Step 2: patterns to grep, useful information in grep results, why you choose this pattern
        - ...
    </reasoning_for_path_choice>
    <conditions_analysis>
        - Detailed breakdown of relevant control-flow structures (if, switch, loops, etc.)
        - Mention explicitly any necessary conditions
        - Explain condition reachability and necessary states to trigger crash
    </conditions_analysis>
    <seed_generation_strategy>
        - Short seed example with explanation for expanding to full seed
        - Python code or pseudocode clearly demonstrating complex seed generation
        - Recommend libraries (e.g., base64, zlib) for accurate and simplified seed generation
    </seed_generation_strategy>
    <script_example>
        # Python script clearly illustrating crash seed generation, the crashing input should be writen into 
          /work/crash.txt using f.write().
    </script_example>
    <conclusion>
        Summarize key findings, critical conditions, and confirm steps to reproduce crash reliably.
    </conclusion>
</report>
\textbf{##Information Provided}:
This project is \textcolor{redtext}{\{project name\}}, in \textcolor{redtext}{\{programming language\}} , you will also receive:
+ Sink function index (<sink_index>), sink function filename (<sink_file_name>), sink function name (<sink_function_name>),
  and sink function code (<sink_function_code>)
+ The harnesses (<harnesses>) code(s) used to reach the sink.
You currently have no information about the call trace. When using the tool, begin your analysis and using the tool by 
treating the sink function as the entry point.
\end{Verbatim}

\end{tcolorbox}
\caption{System Prompt of Code Exploration Agent}
\label{fig:code_agent_prompt}
\end{figure*}

\end{document}